\documentclass[journal]{IEEEtran}
\ifCLASSINFOpdf
\else
\fi
\usepackage{graphicx}
\usepackage{amsmath,amsthm,amssymb}
\usepackage{multirow}
\usepackage{color}
\usepackage{subfig}
\usepackage{soul}
\usepackage{breqn}
\usepackage[table,xcdraw]{xcolor}
\usepackage{stackengine}
\usepackage{tabularx}
\usepackage{todonotes}
\usepackage{booktabs}
\usepackage{xcolor}

\usepackage{soul}
\usepackage{todonotes}

\begin{document}
\title{From Sound Representation to Model Robustness}

\author{Mohammad Esmaeilpour,~\IEEEmembership{Student Member,~IEEE,}
 Patrick Cardinal,~\IEEEmembership{Member,~IEEE,}
 and~Alessandro Lameiras Koerich,~\IEEEmembership{Member,~IEEE}% <-this % stops a space
\thanks{M. Esmaeilpour, P. Cardinal and A. L. Koerich are with the Department
of Software and Information Technology Engineering, \'{E}cole de Technologie Sup\'{e}rieure (\'{E}TS), University of Quebec, Montreal, QC, Canada,
e-mail: (mohammad.esmaeilpour.1@ens.etsmtl.ca, patrick.cardinal@etsmtl.ca, alessandro.koerich@etsmtl.ca).}% <-this % stops a space
%\thanks{Revised manuscript submitted August 19, 2019. Accepted for publication.}}
%\thanks{Accepted for publication November, 2019.}}
}

\markboth{Submitted to IEEE Transaction on audio, speech and language processing,~X, No.~X, January~2021}%
{Esmaeilpour \MakeLowercase{\textit{et al.}}: IEEE Transactions on Information Forensics and Security}

\maketitle

\begin{abstract}
%%%%%%%%%%%%%%%%%%%%%%%%%%%%%%%%%%%%%%%%%%%%%%%%%%%%%%%%%%%%%%%%%%%%
n this paper, we investigate the impact of different standard environmental sound representations (spectrograms) on the recognition performance and adversarial attack robustness of a victim residual convolutional neural network. Averaged over various experiments on three benchmarking environmental sound datasets, we found the ResNet-18 model outperforms other deep learning architectures such as GoogLeNet and AlexNet both in terms of classification accuracy and the number of training parameters. Therefore we set this model as our front-end classifier for subsequent investigations. Herein, we measure the impact of different settings required for generating more informative mel-frequency cepstral coefficient (MFCC), short-time Fourier transform (STFT), and discrete wavelet transform (DWT) representations on our front-end model. This measurement involves comparing the classification performance over the adversarial robustness. On the balance of average budgets allocated by adversary and the cost of attack, we demonstrate an inverse relationship between recognition accuracy and model robustness against six attack algorithms. Moreover, our experimental results show that while the ResNet-18 model trained on DWT spectrograms achieves the highest recognition accuracy, attacking this model is relatively more costly for the adversary compared to other 2D representations.
\end{abstract}

%%%%%%%%%%%%%%%%%%%%%%%%%%%%%%%%%%%%%%%%%%%%%%%%%%%%%%%%%%%%%%%%%%%%

%%%%%%%%%%%%%%%%%%%%%%%%%%%%%%%%%%%%%%%%%%%%%%%%%%%%%%%%%%%%%%%%%%%%

\IEEEpeerreviewmaketitle

%%%%%%%%%%%%%%%%%%%%%%%%%%%%%%%
\section{Introduction}
%%%%%%%%%%%%%%%%%%%%%%%%%%%%%%%

\IEEEPARstart{D}{eveloping} reliable sound recognition algorithms has always been a significant challenge for the signal processing community motivated by real-life applications \cite{marchegiani2017leveraging,salamon2017scaper,radhakrishnan2005audio}. For analyzing surrounding scene either for surveillance \cite{valenzise2007scream} or multimedia sensor networks \cite{steele2013sensor}, there is a constant need for understanding environmental events. Raised by these concerns, several unsupervised \cite{salamon2015unsupervised} and supervised \cite{salamon2017deep} algorithms have been devised for the classification of environmental sounds.

With the proliferation of deep learning (DL) algorithms during the last decade for image-related tasks, large volume of publications on 2D audio representations (spectrograms) have been released. The DL architectures which have been primarily developed for computer vision applications have been well adapted for sound recognition tasks with recognition accuracy competitive to human understanding. However, such algorithms require large amounts of training data. While comprehensive audio datasets are usually not available, many low-level data augmentation approaches have been introduced so that to allow an appropriate training of DL models and improve their performance on sound-related tasks \cite{salamon2017deep}. These approaches apply directly to audio waveforms affecting low-level sampled datapoints of the audio signal which may not necessarily improve performance of the front-end classification models \cite{esmaeilpour2019unsupervised}. In response to this concern, high-level data augmentation approaches have been developed, which are particularly useful for 2D audio representations \cite{kaneko2017generative, mathur2019mic2mic}. Experimental results on a variety of environmental sound datasets attest considerable positive impact of high-level data augmentation on overall performance of DL classifiers (e.g., AlexNet \cite{krizhevsky2012imagenet}, GoogLeNet \cite{szegedy2015going}, etc.) \cite{esmaeilpour2019unsupervised}. A recent study has demonstrated the vulnerability of these convolutional neural networks (ConvNets) trained on 2D representations of audio signals against adversarial attacks \cite{esmaeilpour2019robust}. They have shown that crafted adversarial examples are transferable not only between two dense ConvNets, but also to support vector machines (SVM). This poses a potential harm for sound recognition systems, especially when the highest recognition accuracy has been reported on 2D representations over raw 1D audio signals \cite{boddapati2017classifying}. Developing reliable environmental sound classifiers obliges to study adversarial attacks in greater details and measure their impacts on different sound representations. 

Toward proposing robust classifiers, there have been some debates and case studies on the link between intrusion of adversarial examples and loss functions for some victim classifiers \cite{carlini2017towards}. It has been shown that the integration of more convex loss functions in the victim model (or in the surrogate counterpart) might increase the chance of crafting stronger adversarial examples \cite{carlini2017towards}. It might also depend on some other key factors such as the properties of the classifier, input sample distribution, adversarial setups, etc. In order to study other potential links between robustness and input representation,
%\aktd{links between what? robustness and transferability?}
we evaluate the robustness and the transferability of some state-of-the-art ConvNets against adversarial attacks trained on different 2D sound representations. Our primary front-end ConvNet is ResNet-18 architecture because of its superior recognition performance compared to other ConvNet architectures. We discuss this in Section~\ref{classification_model_Sec} and briefly report our findings on other dense architectures such as GoogLeNet and AlexNet in Section~\ref{discuss:sec}. 

The framework of this paper is environmental sounds encompassing a broad spectrum of the urban sounds, including speech (children playing) and music (street music). However, other specific tasks such as speech recognition, speech-to-text, music genre classification, music recommendation, etc. do not fit in the context of the current paper.
%\aktd{...but there is speech signals and even music in US8k! Need to rewrite this focusing on the task, let's say, excluding specific tasks like we say in the letter}
Our major novelty is investigating classifier response to different representations both in terms of recognition accuracy and robustness against adversarial attacks. This helps to yield more reliable classifiers without running any costly adversarial defense algorithm. More specifically, we make the following contributions:
\begin{itemize}
    \item we show that, models achieve higher recognition accuracy on the DWT representation compared to STFT and MFCC, averaged over different spectrogram settings for three comprehensive environmental sound datasets; 
    \item we identify major spectrogram settings which considerably affect the cost of attack (the number of required gradient computations) averaged over budgets;
    \item we characterize the existence of an inverse relation between recognition accuracy and robustness of the victim models against six strong targeted and non-targeted adversarial attacks. In average, models with higher recognition accuracies undergo higher fooling rates;
    \item we demonstrate that compared to DWT and STFT, the MFCC has relatively lower adversarial transferability ratio among three advanced DL architectures.
\end{itemize}

\noindent The rest of the paper is organized as follows. In Section~\ref{sec:adv}, we briefly review some strong adversarial attacks for 2D audio representations. Explanations on different 2D audio representations that have been used in the experiments are summarized in Section~\ref{audio_rep_section}. Experimental results as well as associated discussions are presented in Section~\ref{sec:exp}. The conclusions and perspectives of future work are presented in the last section. 

%%%%%%%%%%%%%%%%%%%%%%%%%%%%%%%
\section{Adversarial Attacks}
\label{sec:adv}
%%%%%%%%%%%%%%%%%%%%%%%%%%%%%%%
An adversarial attack can be formulated as an optimization problem toward achieving a very small perturbation parameter $\delta$ as stated in Eq.~\ref{general_adv_formula} \cite{szegedy2013intriguing}.
\begin{equation}
 \min_{\delta} \quad f^{*}(\bold{x}+\delta) \neq f^{*}(\bold{x})
 \label{general_adv_formula}
\end{equation}
\noindent where $\bold{x}$ and $f^{*}$ denote a legitimate random sample and the post-activation function of the victim classifier, respectively. The optimal value for $\delta$ should be as small as possible so as to not being perceivable by the human visual system. Although perceiving the applied perturbation on 2D audio representations such as spectrograms is very difficult. To satisfy such a constraint, many attack algorithms have been introduced both in white and black-box scenarios. In this paper, we briefly go over six strong targeted and non-targeted adversarial attacks, which are well adapted to sound recognition models trained on 2D audio representations. We use the average fooling rate of these attacks which is a standard metric for assessing the robustness of victim ConvNets trained on different 2D audio representations.

%%%%%%%%%%%%%%%%%%%%%%%%%%%%%%%
\subsection{Limited-Memory Broyden-Fletcher-Goldfarb-Shanno (L-BFGS)}
%%%%%%%%%%%%%%%%%%%%%%%%%%%%%%%
Szegedy \textit{et al.} \cite{szegedy2013intriguing} discuss that the viability of fooling deep neural networks with fake examples is due to their extremely low probability because such examples are rarely seen in a given dataset. This could be understood as the pitfall of deep networks in low generalizability to unseen but very similar samples. However, they propose an optimization algorithm to mislead finely trained DL models, based on Eq.~\ref{lbfgs}:
\begin{equation}
 \min_{\bold{x}^\prime} c\left \| \delta \right \|_{2} + J_{\bold{w}}(\bold{x}^\prime,l^\prime)
 \label{lbfgs}
\end{equation}
\noindent where $c$ is a positive scaling factor achievable by the line search strategy, 

$\bold{x}^\prime$ denotes the associated crafted adversarial example, $l^\prime$ refers to its target label, and $J_{\bold{w}}$ denotes the loss function for updating weights ($\bold{w}$). There is a variety of choices for this function such as cross-entropy loss or any other surrogate function. The solution of this optimization problem is quite costly and it has been proposed to use %limited-memory Broyden-Fletcher-Goldfarb-Shanno
the L-BFGS optimizer, subject to $0\leq\bold{x}^\prime\leq M$ where $M$ refers to the maximum possible intensity in a spectrogram. This attack is the baseline for the adversarial algorithms that are subsequently presented. 

%%%%%%%%%%%%%%%%%%%%%%%%%%%%%%%
\subsection{Fast Gradient Sign Method (FGSM)}
%%%%%%%%%%%%%%%%%%%%%%%%%%%%%%%
Goodfellow \textit{et al.} \cite{goodfellow2014explaining} explain the existence of adversarial examples of linear nature of deep neural networks even those with super-dense hidden layers. Toward this claim, they proposed a fast optimization algorithm based on Eq.~\ref{fgsm}:
\begin{equation}
 \bold{x}^\prime\leftarrow \bold{x}+\delta \cdot \mathrm{sign}(\nabla_{\bold{x}}J(\bold{x},l))
 \label{fgsm}
\end{equation}
\noindent where $\delta$ is a small constant for controlling the applied perturbation to the legitimate sample $\bold{x}$. Different choices of $\ell_{p}$ norms can be integrated into the FGSM attack, and the adversary should make a trade-off between high similarities and a large enough perturbation to be able to fool 
a model. The formulation of Eq.~\ref{fgsm} for $\ell_{2}$ norm is shown in Eq.~\ref{fgsm2}.
\begin{equation}
 \bold{x}{}'\leftarrow \bold{x}+ \delta \frac{\nabla_{\bold{x}}J(\bold{x},l)}{\left \| \nabla_{\bold{x}}J(\bold{x},l) \right \|}
 \label{fgsm2}
\end{equation}
\noindent where for satisfying the constraint $\bold{x}^\prime\in[0,M]$, the resulting adversarial spectrogram should be clipped or truncated. This white-box adversarial attack is targeted toward a pre-defined wrong label by the adversary in a one-shot scenario.

%%%%%%%%%%%%%%%%%%%%%%%%%%%%%%%
\subsection{Basic Iterative Method (BIM)}
%%%%%%%%%%%%%%%%%%%%%%%%%%%%%%%
This non-targeted adversarial attack \cite{kurakin2016adversarial} is in fact the iterative version of the FGSM optimization algorithm, which crafts and positions potential adversarial examples ideally outside of legitimate subspaces via optimizing Eq.~\ref{bim} for $\delta$:
\begin{equation}
 \bold{x}^\prime_{n+1} \leftarrow \mathrm{clip}_{\bold{x},\delta}\begin{Bmatrix}
\bold{x}^\prime_{n} + \delta \cdot \mathrm{sign} (\nabla_{\bold{x}}J(\bold{x}_{n},l))
\end{Bmatrix}
\label{bim}
\end{equation}
\noindent where $\mathrm{clip}$ is a function for keeping generated examples within the range $[\bold{x}-\delta, \bold{x}+ \delta]$ as defined in Eq.~\ref{clip}.
\begin{equation}
\min \begin{Bmatrix}
M, \bold{x}+\delta, \max\{0,\bold{x}-\delta,\bold{x}^\prime\}
\end{Bmatrix}
\label{clip}
\end{equation}
\noindent where $M$=255 for 8-bit RGB visualization of spectrograms. 

There are two implementations for this optimization algorithm either by keeping optimizing up to reach the first adversarial example (BIM-a) or continue optimizing to a predefined number of iterations (BIM-b). The latter usually generates stronger adversarial examples, though it is more costly since it usually requires more callbacks. Both BIM attacks are iterative and white-box algorithms minimizing Eq.~\ref{bim} for optimal perturbation $\delta$ measured by $\ell_{\infty}$ norm.

%%%%%%%%%%%%%%%%%%%%%%%%%%%%%%%
\subsection{Jacobian-based Saliency Map
Attack (JSMA)}
%%%%%%%%%%%%%%%%%%%%%%%%%%%%%%%
Similar to the FGSM attack, this algorithm also uses gradient information for perturbing the input taking advantage of a greedy approach \cite{papernot2016limitations}. This attack is targeted toward a pre-defined wrong label ($l^\prime$). In fact, it optimizes for $\arg\min_{\delta_{\bold{x}}}\left \| \delta_{\bold{x}} \right \|$ subject to $f^{*}(\bold{x}+\delta_{\bold{x}})=l^\prime$ (optimizing with $\ell_{0}$). There are three steps in developing JSMA adversarial examples. First, computing the derivative of the victim model as Eq.~\ref{jsma_jacob}.
\begin{equation}
 \nabla f(\bold{x})=\frac{\partial f_{j}(\bold{x})}{\partial x_{i}}
 \label{jsma_jacob}
\end{equation}
\noindent where $x_{i}$ denotes pixels intensities. Second, a saliency map should be computed to detect the least effective pixel values for perturbation according to the desired outputs of the model. Specifically, the saliency map for pixels in cases where $\partial f_{l}(\bold{x})/\partial \bold{x}_{i} < 0$ or $\sum_{j\neq l}\partial f_{j}(\bold{x})/\partial \bold{x}_{i}> 0$ should be set to zero since there are detectable variations, otherwise:
\begin{equation}
 S_{map}(\bold{x},l^\prime)[i] = \frac{\partial f_{l}(\bold{x})}{\partial \bold{x}_{i}}\left | \sum_{j\neq l^\prime}\frac{\partial f_{j}(\bold{x})}{\partial \bold{x}_{i}} \right |
 \label{jsma_case2}
\end{equation}
\noindent where $S_{map}$ denotes the saliency map for every given spectrogram $\bold{x}$ and target label $l^\prime$. The last step of the JSMA is applying the perturbation on the original input according to the achieved map.

%%%%%%%%%%%%%%%%%%%%%%%%%%%%%%%
\subsection{Carlini and Wagner Attack (CWA)}
%%%%%%%%%%%%%%%%%%%%%%%%%%%%%%%
This is an iterative and white-box adversarial algorithm \cite{carlini2017towards}, which can use three types of distance metrics: $\ell_{0}$, $\ell_{\infty}$, and $\ell_{2}$ norms. %\aktd{$\ell_{0}$, $\ell_{\infty}$, and $\ell_{2}$ are dissimilarity measures !!!! To be more general I prefer use the term "distance"}
In this paper, we focus on the latter distance measure, which makes the algorithm very strong even against distillation network. The optimization problem in this attack is given by Eq.~\ref{cwa_minim}.
\begin{equation}
 \min_{\delta}\left \| \bold{x}^\prime- \bold{x} \right \|_{2}^{2}+cf(\bold{x}^\prime)
 \label{cwa_minim}
\end{equation}
\noindent where $c$ is a constant value as explained in Eq.~\ref{lbfgs}. Assuming the target class is $l^\prime$ and $G(\bold{x}^\prime)_{i}$ denotes the logits of the trained model $f$ before softmax activation corresponding to the $i$-th class, then:

\begin{equation}
 f(\bold{x}^\prime)=\max \left\{ \max_{i\neq l^\prime}\left \{ G(\bold{x}^\prime)_{i}\right \} - G(\bold{x}^\prime)_{l^\prime}, -\kappa \right \}
 \label{cwa_func}
\end{equation}
%}
\noindent where $\kappa$ is a tunable confidence parameter for increasing misclassification confidence toward label $l^\prime$. The actual adversarial example is given by Eq.~\ref{cwa_advEx}.
\begin{equation}
 \bold{x}'=\frac{1}{2}\left [ \tanh(\mathrm{arctanh}(\bold{x})+\delta)+1 \right ]
 \label{cwa_advEx}
\end{equation}
\noindent where the $\tanh$ activation function is used in replacement of box-constraint optimization. For non-targeted attacks, Eq.~\ref{cwa_func} should be updated as: 

\begin{equation}
f(\bold{x}^\prime) = \max \left\{ G(\bold{x}^\prime)_{l}-\max_{i\neq l}\left \{ G(\bold{x}^\prime)_{i} \right \} ,-\kappa\right\}
\end{equation}

%%%%%%%%%%%%%%%%%%%%%%%%%%%%%%%
\subsection{DeepFool Adversarial Attack}
%%%%%%%%%%%%%%%%%%%%%%%%%%%%%%%
Moosavi-Dezfooli \textit{et al.} \cite{moosavi2016deepfool} proposed a white-box algorithm for finding the most optimal perturbation for redirecting the position of a legitimate sample toward a pre-defined target label using linear approximation. The optimization problem for achieving optimal $\delta$ is given by Eq.~\ref{deepfool_eq1}.
\begin{equation}
 \arg \min \left \| \delta \right \|_{2} \quad \mathrm{s.t.} \quad \mathrm{sign}(f(\bold{x}^\prime))\neq \mathrm{sign}(f(\bold{x}))
 \label{deepfool_eq1}
\end{equation}
\noindent where $\delta = -f(\bold{x})\bold{w}/\left \| \bold{w} \right \|_{2}^{2}$. DeepFool can also be modified to a non-targeted attack optimizing for hyperplanes of the victim model. In this paper, we implement targeted DeepFool attack and averaged over available labels measuring over $\ell_{2}$ and $\ell_{\infty}$. In practice, this scenario is not only faster but also more destructive than BIMs. 

In the next section we provide a brief overview of common 2D representations of audio signals using time-frequency transformations. Finally, we carry out our adversarial experiments on the transformed audio signals (spectrograms).

%%%%%%%%%%%%%%%%%%%%%%%%%%%%%%%%%%%%%%%%%%%%%%%%%%
\section{2D Audio Representations}
\label{audio_rep_section}
%%%%%%%%%%%%%%%%%%%%%%%%%%%%%%%%%%%%%%%%%%%%%%%%%%
Representing audio signals using time-frequency plots is a standard operation in audio and speech processing, which aims representing such signals into a compact and informative way. Fourier and wavelet transforms are the most commonly used approaches for mapping an audio signal into frequency-magnitude representations. In this section we briefly review some of the most common approaches.

%%%%%%%%%%%%%%%%%%%%%%%%%%%%%%%
\subsection{Short-Time Fourier Transform (STFT)}
%%%%%%%%%%%%%%%%%%%%%%%%%%%%%%%
For a given continuous signal $a(t)$ which is distributed over time, its STFT using a Hann window function $w(\tau)$ can be computed using Eq.~\ref{STFT_contin}.
\begin{equation}
 \mathrm{STFT}\begin{Bmatrix} a(t) \end{Bmatrix}(\tau, \omega)=\int_{-\infty}^{\infty}a(t)w(t-\tau)e^{-j\omega t}dt
 \label{STFT_contin}
\end{equation}
\noindent where $\tau$ and $\omega$ are time and frequency axes, respectively. This transform is quite generalizable to discrete time domain for a discrete signal $a[n]$ as:
\begin{equation}
 \mathrm{STFT}\begin{Bmatrix} a[n] \end{Bmatrix}(m,\omega)=\sum_{n=-\infty}^{\infty}a[n]w[n-m]e^{-j\omega n}
 \label{STFT_disc}
\end{equation}
\noindent where $m\ll n$ and $\omega$ is a continuous frequency coefficient. In other words, for generating the STFT of a discrete signal, we need to divide it into overlapping shorter length sub-signals and compute Fourier transform on it, which results in an array of complex coefficients. Computing the square of the magnitude of this array yields to a spectrogram representation as shown in Eq.~\ref{STFT_spec}.
\begin{dmath}
 \mathrm{Sp_{STFT}}\begin{Bmatrix} a[n] \end{Bmatrix}(m,\omega)=
 \left | \sum_{n=-\infty}^{\infty}a[n]w[n-m]e^{-j\omega n} \right |^2
 \label{STFT_spec}
\end{dmath}
\noindent This 2D representation shows frequency distribution over discrete time and compared to the original signal $a[n]$, it has a lower dimensionality, although it is a lossy operation.

%%%%%%%%%%%%%%%%%%%%%%%%%%%%%%%
\subsection{Mel-Frequency Cepstral Coefficients (MFCC)}
%%%%%%%%%%%%%%%%%%%%%%%%%%%%%%%
This transform is a variation of the STFT with some additional postprocessing operations including non-linear transformation. For every column of the achieved spectrogram, we compute its dot product with a number of Mel filter bank (power estimates of amplitudes distributed over frequency). For increasing the resolution of the resulting vector, logarithmic filtering should be applied and finally, it will be mapped to another 1D representation using discrete cosine transform.

This representation has been widely used for sound enhancement and classification.
Furthermore, it has been well studied as a standard approach for conventional generative models incorporating Markov chain and Gaussian mixture modes \cite{shi2018hidden, maurya2018speaker}.

%%%%%%%%%%%%%%%%%%%%%%%%%%%%%%%
\subsection{Discrete Wavelet Transform (DWT)}
%%%%%%%%%%%%%%%%%%%%%%%%%%%%%%%
Wavelet transform maps the continuous signal $a(t)$ into time and scale (frequency) coefficients similar to STFT using Eq.~\ref{dwt_contin}. 
\begin{equation}
 \mathrm{DWT}\begin{Bmatrix} a(t) \end{Bmatrix} = \frac{1}{\sqrt{\left | s \right |}}\int_{-\infty}^{\infty}a(t)\psi \begin{pmatrix} \frac{t-\tau}{s} \end{pmatrix}dt
 \label{dwt_contin}
\end{equation}
\noindent where $s$ and $\tau$ denote discrete scale and time variations, respectively, and $\psi$ is the core transformation function

which is also known as mother function (see Eq.~\ref{morlet_func}). There are a variety of mother functions for different applications such as the complex Morlet which is given by Eq.~\ref{morlet_func}: 

\begin{equation}
 \psi(t)=\frac{1}{\sqrt{2\pi}}e^{-j\omega t}e^{-t^{2}/2}
 \label{morlet_func}
\end{equation}
\noindent Discrete time formulation for this transform is shown in Eq.~\ref{dwt_discrete}.
\begin{equation}
 \mathrm{DWT}\begin{Bmatrix} a[k,n] \end{Bmatrix}=\int_{-\infty}^{\infty}a(t)h\begin{pmatrix} na^{k}T-t \end{pmatrix}
 \label{dwt_discrete}
\end{equation}
\noindent where $n$ and $k$ are integer values for the continuous mother function of $h$. Spectral representation for this transformed signal is a 2D array which is computed by Eq.~\ref{dwt_spectrogram}:
\begin{equation}
 \mathrm{Sp_{DWT}}\begin{Bmatrix} a(t) \end{Bmatrix}=\left | \mathrm{DWT}\begin{Bmatrix} a[k,n] \end{Bmatrix} \right |
 \label{dwt_spectrogram}
\end{equation}

In the next section, we explain our experiments on three benchmarking sound datasets. We firstly generate separate spectrogram sets with the three aforementioned representation using different configurations. Second, we train a ResNet on these datasets and run adversarial attack algorithms against them. Finally, we measure both the fooling rate and the cost of attacks. We demonstrate that for different spectrogram configurations these metrics are meaningfully different.  

%%%%%%%%%%%%%%%%%%%%%%%%%%%%%%%
\section{Experiments}
\label{sec:exp}
%%%%%%%%%%%%%%%%%%%%%%%%%%%%%%%
We use three environmental sound datasets in all our experiments: UrbanSound8k \cite{Salamon:UrbanSound:ACMMM:14}, ESC-50 \cite{piczak2015esc}, and ESC-10 \cite{piczak2015esc}. The first dataset includes 8732 four-second length audio samples distributed in 10 classes: engine idling, car horn, children playing, drilling, air conditioner, jackhammer, dog bark, siren, gun shot, and street music. ESC-50 is a comprehensive dataset with 50 different classes and overall 2000 five-second length audio recordings of natural acoustic sounds. A subset of this dataset is ESC-10 which has been released with 10 classes and 400 recordings. 

For increasing both the quality and the quantity of samples of these datasets we apply pitch shifting augmentation approach with scales $0.75$, $0.9$, $1.15$, and $1.5$ as proposed in \cite{esmaeilpour2019unsupervised}, which positively affect classification accuracy. This data augmentation operation generates four extra audio samples for every original audio sample and eventually it increases the size of the original dataset by the factor of four. We discuss the usefulness of this 1D data augmentation approach in Section~\ref{discuss:sec}. In the following subsection, we explain the details of generating 2D representations for audio signals. To this aim we use the open-source Librosa signal processing python library \cite{mcfee2015librosa} and our upgraded version of the wavelet toolbox \cite{waveletSoundExplorer}.

%%%%%%%%%%%%%%%%%%%%%%%%%%%%%%%
\subsection{Generating Spectrograms}
%%%%%%%%%%%%%%%%%%%%%%%%%%%%%%%
For every dataset including augmented signals we separately generate independent sets of 2D representations, namely MFCC, STFT, and DWT. Our aim is to investigate which audio representation yields a better trade-off between recognition accuracy and robustness for a victim model against a variety of strong adversarial attacks.

%%%%%%%%%%%%%%%%%%%%%%%%%%%%%%%
\subsubsection{MFCC Production Settings}
%%%%%%%%%%%%%%%%%%%%%%%%%%%%%%%
There are four major settings in generating MFCC spectrogram using Librosa. The default value for sampling rate is 22050 Hz. Since there is no optimal approach for determining the best sampling rate so that generate the most informative spectrogram, we run extensive experiments using sampling rates from 8 to 24 kHz. The second tunable hyperparameter is the number of MFCCs ($N_{\text{MFCC}}$) which we examine different values for it: 13, 20, and 40 per frame with hop length of 1024. Normalization of discrete cosine transform (type 2 or 3) using orthonormal DCT basis for MFCC production is the third setting. By default, this hyperparameter is set to true in almost all the libraries including Librosa, although we measure performance of the front-end classifier trained to MFCC spectrograms without normalization. The last argument is about the number of cepstral filtering ($CF$) \cite{juang1987use} to be applied on MFCC features. The sinusoidal $CF$ reduces involvement of higher order coefficients and improve recognition performance \cite{paliwal1999decorrelated} (see Eq.~\ref{lifter_cf}).
\begin{equation}
 \bold{M}\left [ n,: \right ] \leftarrow \bold{M}\left [ n,: \right ] \times \left ( 1+\sin \left( \frac{\pi(n+1)}{CF} \right) \right ) \frac{CF}{2}
 \label{lifter_cf}
\end{equation}
\noindent where $\bold{M}$ stands for MFCC array with size $[n,:]$. We investigate the effect of $CF$ on the overall performance of classification models.

%%%%%%%%%%%%%%%%%%%%%%%%%%%%%%%
\subsubsection{STFT Production Settings}
%%%%%%%%%%%%%%%%%%%%%%%%%%%%%%%
For producing STFT representations, we use default configurations for general hyperparameters as outlined in the Librosa manual. For assigning the length of the windowed signal, we use 2048, 1024, and 512 with associated sampling rates. We also use variable window size: 2048 (default value), 1024, and 512 (very small window) associated with default hop size of 512. We investigate potential effects of these configurations for the resiliency of the victim models against adversarial attacks.

%%%%%%%%%%%%%%%%%%%%%%%%%%%%%%%
\subsubsection{DWT Production Settings}
%%%%%%%%%%%%%%%%%%%%%%%%%%%%%%%
For generating DWT representations, we modified the sound explorer software \cite{waveletSoundExplorer} to support Haar and Mexican Hat wavelet mother functions in addition to complex Morlet. Sampling frequency for DWT spectrograms has been set up to 8 kHz and 16 kHz with constant frame length of 50 ms. Moreover, by convention, the overlapping threshold is set to 50\%. In our experiments we measure the impacts of these DWT configurations visualized in logarithmic scale (for higher resolution) on both recognition accuracy and robustness against adversarial attacks. 

In the next subsection, we discuss possible choices for the classification models to be separately trained on the aforementioned spectrogram representations and setups. We select our final front-end classifier from a diverse domain of traditional handcrafted-based feature learning algorithms to state-of-the-art DL architectures.

%%%%%%%%%%%%%%%%%%%%%%%%%%%%%%%
\subsection{Classification Model}
\label{classification_model_Sec}
%%%%%%%%%%%%%%%%%%%%%%%%%%%%%%%
For the choice of classification algorithms, we initially included both conventional classifiers such as linear and Gaussian SVM \cite{esmaeilpour2019robust}, random forest \cite{esmaeilpour2019unsupervised}, and some deep leaning architectures. Specifically, we selected pre-trained GoogLeNet, AlexNet, and ResNet \cite{he2016deep} models tuned for our three benchmarking datasets. We preserved the architectures of these ConvNets except for the first layer and the last layer for mapping logits into class labels (softmax layer). Since spectrograms may have different dimensions according to their length and transformation schemes, we bilinearly interpolate them to fit 128$\times$128 for all the ConvNets.

Performance comparison of the aforementioned SVMs, GoogLeNet and AlexNet against a few adversarial attacks have already been studied mainly for DWT representations of environmental sound datasets in \cite{esmaeilpour2019robust}. Their experiments have been conducted on standard spectrograms without validating potential impacts of different settings in the process of producing different representations. In this paper, we carry out extensive experiments using: (i) three common 2D representations for audio signals, namely MFCC (represented in 2D matrix format not the common vector visualization), STFT, and DWT; (ii) more and stronger targeted and non-targeted algorithms for adversarial attacks; (iii) fair comparison on fooling rates of victim models taking their cost of attacks averaged over the allocated budgets into account. 

We primarily select a ConvNet as our front-end classifier for the sake of simplicity and interpretability of results. We present concise results for other classification models in Section~\ref{discuss:sec}. For such an aim, we selected ResNet architectures because such ConvNet is currently the best-performing classifiers for several tasks \cite{hershey2017cnn}. Our implementations corroborate that, on average, these ConvNet architectures outperform all the above-mentioned algorithms (both SVMs and other DL approaches) trained on spectrograms. Among the possible architectures for ResNet (ResNet-18, ResNet-34, and ResNet-56), we selected ResNet-18 according to its highest recognition performance and relatively low number of parameters compared to others. Recalling that, we investigate potential effects of spectrogram configurations on the classifier which not only has a very competitive recognition accuracy compared to others, but also requires less number of training parameters. Thus herein, we specifically focus on the ResNet-18 network and all our investigations will consider this victim architecture.

For every configuration to produce the 2D representations, we generate an individual set of spectrograms and train an independent ResNet-18 classifier on each set. We use a 5-fold cross validation setup on 70\% of the overall dataset volume (training plus development). For avoiding overtraining, we implemented early stopping technique in training and finally report mean recognition accuracy on the test sets (30\% remaining). 

%%%%%%%%%%%%%%%%%%%%%%%%%%%%%%%
\subsection{Adversarial Attacks}
\label{adv:setup}
%%%%%%%%%%%%%%%%%%%%%%%%%%%%%%%
In this section, we provide details for attacking the models trained on 2D audio representations. We examine their robustness against six strong adversarial attacks by reporting obtained average fooling rates using the area under the ROC curve (AUC) as a performance metric.

%%%%%%%%%%%%%%%%%%%%%%%%%%%%%%%
\subsubsection{Settings for Attack Algorithms}
%%%%%%%%%%%%%%%%%%%%%%%%%%%%%%%
In FGSM and BIMs attacks, possible ranges for $\delta$ have been defined from $0.001$ to any possible supremum under different confidence intervals ($\geq 65\%$). For the implementation of the DeepFool attack, we use the open-source Foolbox package \cite{rauber2017foolbox} with iterations from 100 to 1000 (10 different scales with a step of 100). In the implementation of the JSMA attack, the number of iterations has been set to $(m_{i} \gamma)/n_{i}$ where $m_{i}$ and $n_{i}$ denote the total number of pixels and scaling factor within $[0, 200]$ (with displacement a of 40), respectively. Also $\gamma$ is the maximum allowed distortion (ideally $< 1.5/255$) within the maximum number of iterations. Budget allocated to CWA is within $\left \{1, 3, 7, 9 \right \}$ for search steps in $c$ within $\left \{25, 100, 1\mathrm{k}, 2\mathrm{k}, 5\mathrm{k} \right \}$ iterations in each search step using early stopping. For targeted attacks (i.e., FGSM, JSMA, and CWA) we randomly select targeted wrong labels for running adversarial optimization algorithms.

We executed these attack algorithms on two NVIDIA GTX-1080-Ti with $4 \times 11$ GB of memory except for the DeepFool attack, which was executed on 64-bit Intel Core-i7-7700 (3.6 GHz) CPU with 64 GB memory. For attacks on the smallest dataset (ESC-10), we used batches of 200 samples. For larger datasets (ESC-50 and UrbanSound8k), we used 25 batches of 100 samples.

%%%%%%%%%%%%%%%%%%%%%%%%%%%%%%%
\subsubsection{Adversarial Attacks for MFCC Representations}
%%%%%%%%%%%%%%%%%%%%%%%%%%%%%%%
We firstly investigate the potential effect of different sampling rates in MFCC production on the performance of the trained models. To this end, sampling rates have been selected from fairly low (8 kHz) to moderately high (24 kHz) ranges including the default frequency value (22.05 kHz) defined in Librosa. Therefore, we trained four ResNet-18 models per dataset associated with four sampling rates. The results summarized in Table~\ref{mfcc_sr_effect} show that the recognition performance of the classifiers is, to some extent, dependent on the sampling rates. For ESC-10 and UrbanSound8k datasets, the sampling rate of 8 kHz improves recognition accuracy while 16 kHz works better for ESC-50. These results might imply that a low sampling rate filters out high frequency components and negatively affects learning of discriminative features from the spectrograms. 

\begin{table*}[htpb]
\centering
\caption{Performance comparison of models trained on MFCC representations with different sampling rates averaged over experiments and budgets. Relatively better performances are in bold face.}
\begin{tabular}{|c|c|c||c|c|c|c|c|c|}
\hline
\multirow{2}{*}{Dataset} & Sampling & Recognition & \multicolumn{6}{c|}{AUC Score, Number of Gradients for Adversarial Attacks} \\ \cline{4-9} 
 & Rate (kHz) & Accuracy (\%) & FGSM & DeepFool & BIM-a & BIM-b & JSMA & CWA \\ \hline
\multirow{4}{*}{ESC-10} & 8 & \textbf{73.23} & \textbf{0.9822}, 1 & 0.9473, 074 & \textbf{0.9710}, 065 & \textbf{0.9801}, 110 & \textbf{0.9308}, 096 & \textbf{0.9912}, 1346 \\ \cline{2-9} 
 & 16 & 72.15 & 0.9456, 1 & \textbf{0.9607}, 046 & 0.9334, 059 & 0.9375, 197 & 0.9144, 151 & 0.9616, 1435 \\ \cline{2-9} 
 & 22.05 & 72.06 & 0.9467, 1 & 0.9518, 129 & 0.9309, 088 & 0.9379, 186 & 0.9145, 213 & 0.9405, 1471 \\ \cline{2-9} 
 & 24 & 70.13 & 0.9471, 1 & 0.9341, 078 & 0.9298, 115 & 0.9327, 171 & 0.9233, 091 & 0.9302, 1149 \\ \hline \hline
\multirow{4}{*}{ESC-50} & 8 & 69.89 & 0.9517, 1 & 0.9023, 061 & 0.9612, 084 & 0.9703, 193 & 0.9288, 118 & 0.9598, 2418 \\ \cline{2-9} 
 & 16 & \textbf{70.21} & \textbf{0.9849}, 1 & \textbf{0.9912}, 248 & \textbf{0.9871}, 209 & \textbf{0.9903}, 160 & \textbf{0.9508}, 251 & 0.9672, 2639 \\ \cline{2-9} 
 & 22.05 & 69.97 & 0.9534, 1  & 0.9386, 331 & 0.9430, 423 & 0.9581, 288 & 0.9233, 219 & 0.9434, 2318 \\ \cline{2-9} 
 & 24 & 67.25 & 0.9433, 1 & 0.9214, 208 & 0.9307, 159 & 0.9415, 216 & 0.9187, 417 & \textbf{0.9652}, 2744 \\ \hline \hline
\multirow{4}{*}{UrbanSound8k} & 8 & \textbf{71.25} & \textbf{0.9905}, 1 & \textbf{0.9895}, 326 & 0.9411, 317 & \textbf{0.9950}, 223 & \textbf{0.9623}, 398 & \textbf{0.9708}, 2791 \\ \cline{2-9} 
 & 16 & 70.81 & 0.9508, 1 & 0.9215, 631 & 0.9346, 519 & 0.9389, 817 & 0.9447, 442 & 0.9449, 3805 \\ \cline{2-9} 
 & 22.05 & 69.57 & 0.9457, 1 & 0.9151, 269 & \textbf{0.9449}, 184 & 0.9256, 513 & 0.9370, 416 & 0.9456, 3015 \\ \cline{2-9} 
 & 24 & 69.33 & 0.9440, 1 & 0.9221, 318 & 0.9236, 299 & 0.9120, 862 & 0.9242, 343 & 0.9371, 2816 \\ \hline
\end{tabular}
\label{mfcc_sr_effect}
\end{table*}

We attack these models using the aforementioned six adversarial algorithms and measure their fooling rates averaged over different budgets as explained in Section \ref{adv:setup}. From the results shown in Table~\ref{mfcc_sr_effect}, we notice an inverse relationship between recognition accuracy and robustness of these models, on average. For instance, ResNet-18 trained on MFCC spectrograms of the ESC-10 dataset sampled at 8 kHz reaches the highest recognition accuracy, but this model is less robust against five out of six adversarial attacks, averaged over the allocated budgets. We present two hypotheses on this issue. Firstly, adversarial attacks are essentially optimization-based problems and their final results are dependent on the hyperparameters defined by the adversary. Confidence intervals, number of callbacks to the original spectrogram, number of iterations in optimization formulation, line search for the optimal coefficient are among those to name a few. Fooling rate of a victim model is dependent on tuning these hyperparameters. Our second hypothesis is on the statistical perspective of training a neural network. A model with higher recognition accuracy has probably learned a better decision boundary via maximizing the intra-class similarity and inter-class dissimilarity. Attacking this model, provides a wider search area for the adversary to find pinholes of the model, especially when the decision boundaries among classes lie in the vicinity of each other. Table~\ref{mfcc_sr_effect} also compares average number of gradient for batch execution required by every attack algorithm. Regarding statistics of this table, CWA is the costliest adversarial attacks for spectrograms with different sampling rates.

\begin{figure}[htpb!]
  \centering
  \includegraphics[width=0.5\textwidth]{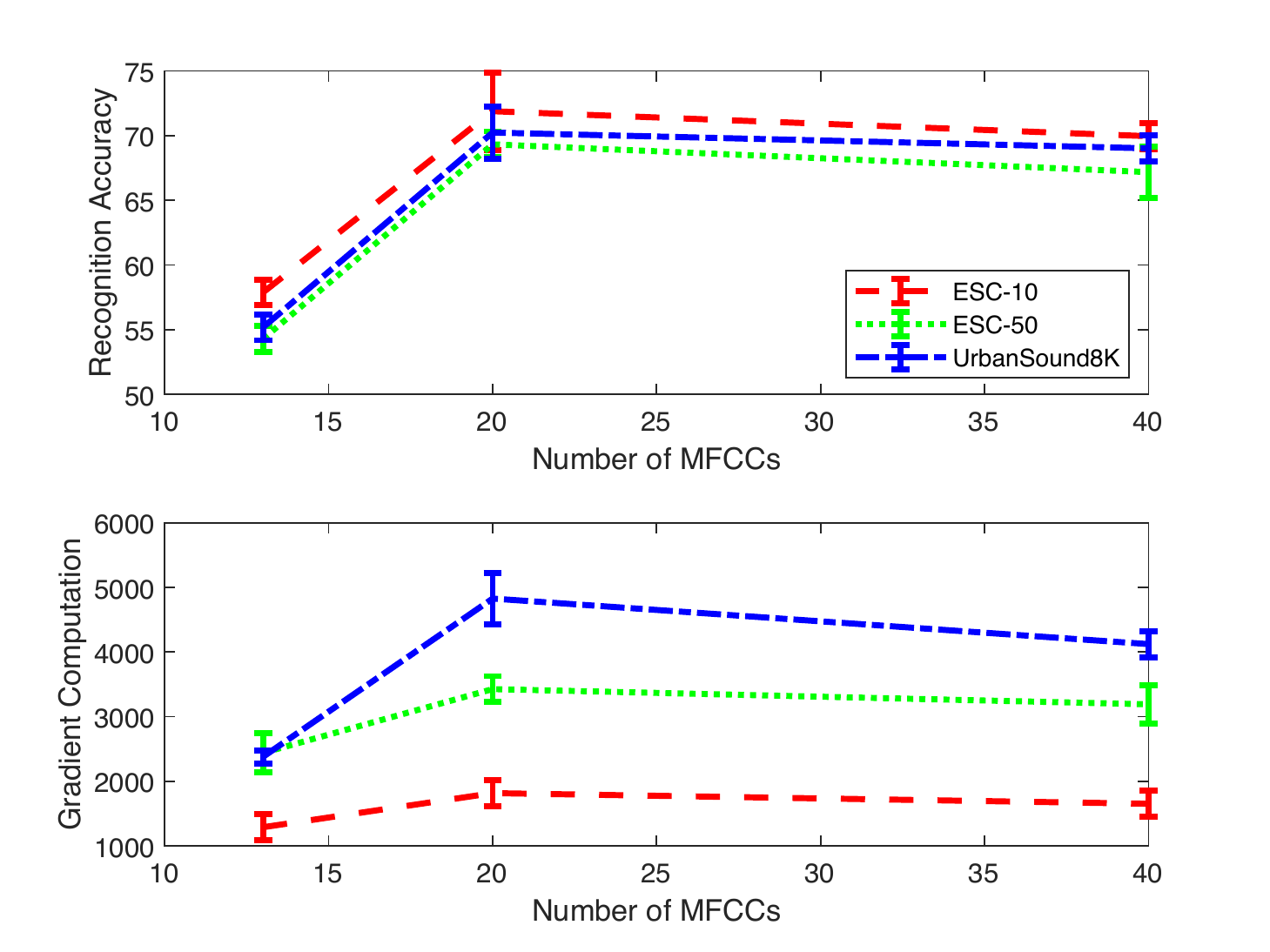}
  \caption{Effect of $N_{\mathrm{MFCC}}$ on the recognition accuracy and on the average cost of the attack (number of batch gradient computation) over six adversarial algorithms for ResNet-18 models.}
  \label{mel-attack-pack}
\end{figure}

The default value for the number of MFCCs ($N_{\text{MFCC}}$) is 20 as defined in Librosa. However, we encompass values from a minimum number of 13 to a maximum of 40 in generating MFCC spectrograms; although increasing $N_{\text{MFCC}}$$>$20 introduces redundancy in frequency coefficient representation. Our experimental results corroborate the negative effect of a low $N_{\text{MFCC}}$ in the performance of the classifiers. More specifically, recognition performance of the trained models on spectrograms with $N_{\text{MFCC}}=$ 13 is 14\% less than models trained on spectrograms with $N_{\text{MFCC}}$$\geq$20, on average. Our experimental results on attacking victim models trained on spectrograms with low $N_{\text{MFCC}}$ unveils their extreme vulnerabilities. However, in terms of cost of the attack, these models need fewer callbacks for gradient computations for yielding AUC$>$90\%  (see Fig.
~\ref{mel-attack-pack}). This could be due to the nature of adversarial attacks as optimization formulations regardless of the performance of the victim models.

\begin{figure}[htpb!]
  \centering
  \includegraphics[width=0.5\textwidth]{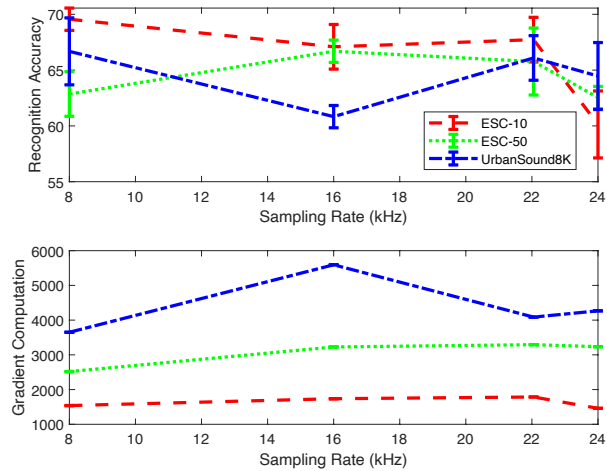}
  \caption{Normalization effect on the recognition accuracy and on the average cost for reaching AUC$>$0.9 over six adversarial algorithms for ResNet-18 models. }
  \label{mel-attack-pack2}
\end{figure}

Using orthonormal discrete cosine transform basis function is a standard approach in crafting MFCC spectrograms. In our experiments we produced two separate subsets of spectrograms with and without normalization to measure its potential effect on both the recognition accuracy and the fooling rate (see Fig.~\ref{mel-attack-pack2}). Disabling this normalization scheme causes a drop in the recognition accuracy and in the cost of the attack of the models to 7\% and 8.5\%, respectively on average. 

\begin{figure}[htpb!]
  \centering
  \includegraphics[width=0.5\textwidth]{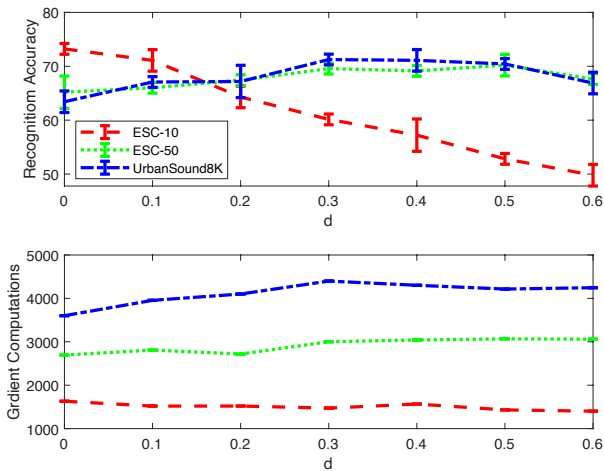}
  \caption{Cepstral filtering effect on the recognition accuracy and on the average cost for reaching AUC$>$0.9 of attack over six adversarial algorithms for ResNet-18 models.}
  \label{mel-attack-pack3}
\end{figure}

For the choice of the cepstral filtering, we covered values in the range $\begin{bmatrix} 0, \left (d \times N_{\text{MFCC}} \right ) \end{bmatrix}$ where maximum $d$ is 2.5 with hop size of 0.5 in the production of spectrograms. Values above the supremum of this interval generate higher-order coefficients in linear-like weighting distributions which considerably reduce recognition accuracy on average to about 48\%. Optimal values for $d$ are 0, 0.5, and 0.3 for ESC-10, ESC-50, and UrbanSound8k, respectively (see Fig.~\ref{mel-attack-pack3}).

%%%%%%%%%%%%%%%%%%%%%%%%%%%%%%%
\subsubsection{Adversarial Attacks for STFT Representations}
%%%%%%%%%%%%%%%%%%%%%%%%%%%%%%%
There is a significant similarity in producing MFCC and STFT spectrograms mainly in terms of transformation and frequency modulation. Therefore, we omit experimental results relevant to measuring impacts of sampling rates on the robustness of victim classifiers. Fooling rates of ResNet-18 models on STFT representations are similar to MFCC representations and such rates support the aforementioned inverse relationship between recognition accuracy and robustness against attacks.

\begin{figure}[htpb!]
  \centering
  \includegraphics[width=0.5\textwidth]{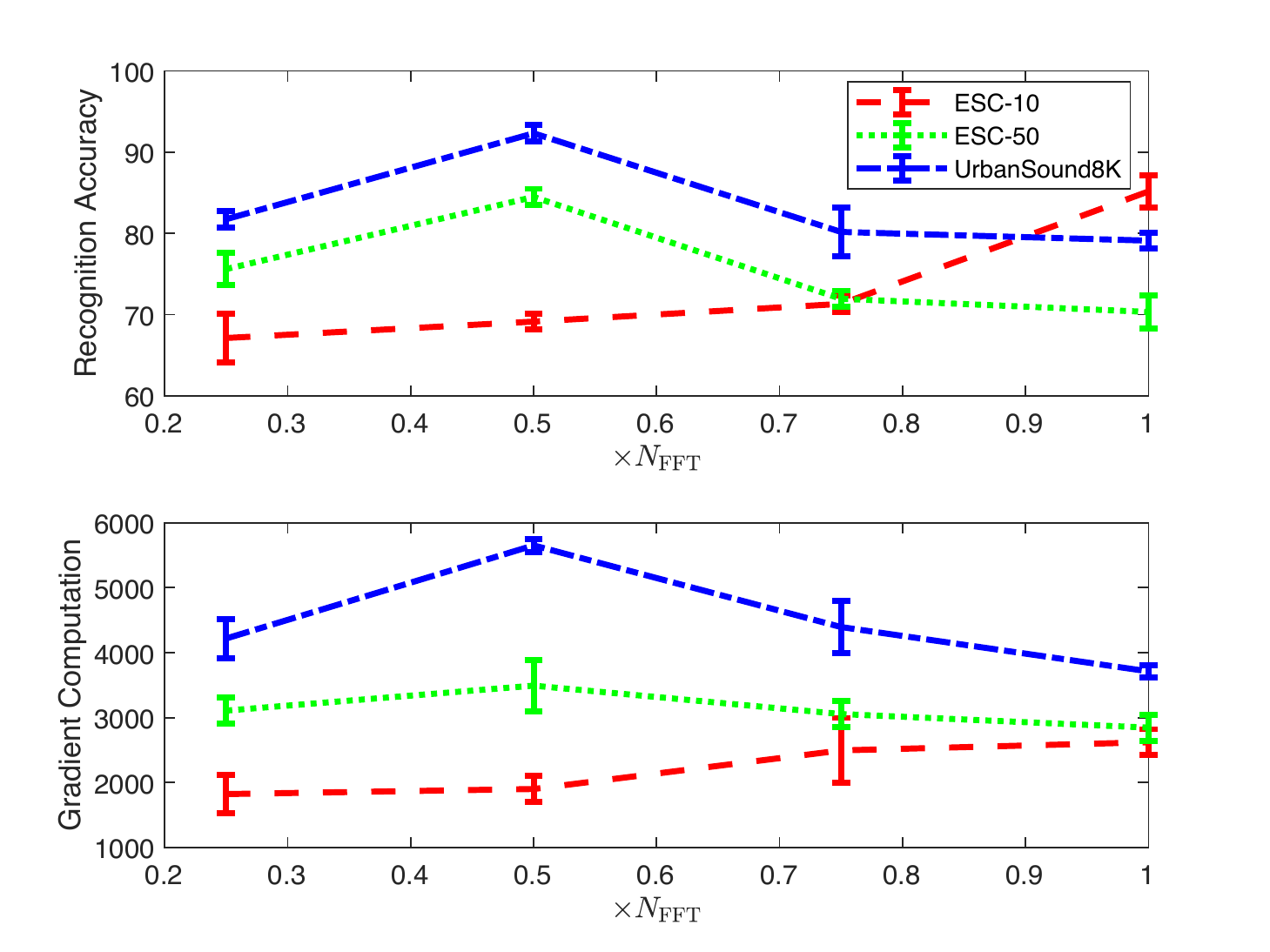}
  \caption{Effect of scales for $N_{\mathrm{FFT}}$ on the recognition accuracy and on the average cost of the attack for reaching AUC$>$0.9 over six adversarial algorithms for ResNet-18 models.}
  \label{mel-attack-pack4}
\end{figure}

Table~\ref{stft_fft_effect} summarizes adversarial experiments conducted on STFT representations with the same aforementioned setup described in Section~\ref{adv:setup}. This table illustrates the impact of the number of FFTs ($N_{\text{FFT}}$) both on the recognition accuracy and on the robustness of victim models against adversarial attacks averaged over all the different adversarial setups. For ESC-10 and ESC-50 datasets, $N_{\text{FFT}}$$=$1024 results in learning better decision boundaries for the classifiers, although it increases fooling rates of the victim models. In production of STFT spectrograms, each frame of a given audio signal is spanned by a window which covers the frame. The maximum length of this window can be equivalent to the number of $N_{\text{FFT}}$. Since small window lengths improve the temporal resolution of the final STFT representation, we evaluate the performance of the models on small window lengths in the range $\begin{bmatrix}\begin{pmatrix}0.25$$\times$$N_{\text{FFT}}\end{pmatrix}, N_{\text{FFT}}\end{bmatrix}$ with hop size of $N_{\text{FFT}}/4$. As shown in Fig.~\ref{mel-attack-pack4}, the evaluation on ESC-50 and UrbanSound8k datasets uncovers that models trained on STFT representations with window length of $0.5$$\times$$N_{\text{FFT}}$ outperform others. On the ESC-10 dataset, window length of $N_{\text{FFT}}$ resulted in better performance in terms of recognition accuracy.
%windowsize-std.eps

\begin{table*}[htpb!]
\centering
\caption{Performance comparison of models trained on STFT representations with different $N_{\text{FFT}}$ averaged over experiments and budgets. Relatively better performances are in bold face.}
\begin{tabular}{|c|c|c||c|c|c|c|c|c|}
\hline
\multirow{2}{*}{Dataset} & Number & Recognition & \multicolumn{6}{c|}{AUC Score, Number of Gradients for Adversarial Attacks} \\ \cline{4-9} 
 & of FFTs & Accuracy (\%) & FGSM & DeepFool & BIM-a & BIM-b & JSMA & CWA \\ \hline
\multirow{3}{*}{ESC-10} & 512 & 82.41 & 0.9768, 1 & 0.9430, 089 & 0.9576, 109 & 0.9717, 134 & 0.9662, 141 & 0.9846, 1415 \\ \cline{2-9} 
 & 1~024 & \textbf{85.17} & \textbf{0.9823}, 1 & \textbf{0.9701}, 129 & \textbf{0.9715}, 091 & \textbf{0.9792}, 183 & 0.9531, 209 & \textbf{0.9905}, 2008 \\ \cline{2-9} 
 & 2~048 & 80.56 & 0.9651, 1 & 0.9544, 092 & 0.9407, 163 & 0.9529, 279 & 0.9588, 341 & 0.8731, 1730 \\ \hline \hline
\multirow{3}{*}{ESC-50} & 512 & 82.44 & 0.9786, 1 & 0.9542, 082 & 0.9583, 109 & 0.9665, 244 & 0.9614, 128 & 0.9618, 1995 \\ \cline{2-9} 
 & 1~024 & \textbf{84.49} & \textbf{0.9881}, 1 & 0.9512, 331 & \textbf{0.9871}, 267 & \textbf{0.9798}, 179 & 0.9702, 361 & \textbf{0.9896}, 2353 \\ \cline{2-9} 
 & 2~048 & 83.12 & 0.9567, 1 & \textbf{0.9631}, 145 & 0.9765, 211 & 0.9606, 567 & \textbf{0.9738}, 399 & 0.9729, 2412 \\ \hline \hline
\multirow{3}{*}{UrbanSound8k} & 512 & 90.58 & 0.9761, 1 & 0.9414, 583 & \textbf{0.9513}, 442 & 0.9682, 421 & 0.9402, 345 & 0.9539, 2569 \\ \cline{2-9} 
 & 1~024 & 91.74 & 0.9827, 1 & 0.9752, 322 & 0.9340, 471 & \textbf{0.9687}, 719 & 0.9515, 502 & 0.9654, 3271 \\ \cline{2-9} 
 & 2~048 & \textbf{92.23} & \textbf{0.9895}, 1 & \textbf{0.9764}, 643 & 0.9407, 602 & 0.9630, 408 & \textbf{0.9623}, 655 & \textbf{0.9673}, 3342 \\ \hline
\end{tabular}
\label{stft_fft_effect}
\end{table*}

\begin{table*}[htpb!]
\centering
\caption{Performance comparison of models trained on DWT representations with different sampling rates averaged over different budgets. Relatively better performances are shown in bold face.}
\begin{tabular}{|c|c|c||c|c|c|c|c|c|}
\hline
\multirow{2}{*}{Dataset} & \multirow{2}{*}{\begin{tabular}[c]{@{}c@{}}Sampling\\ Rate (kHz)\end{tabular}} & \multirow{2}{*}{\begin{tabular}[c]{@{}c@{}}Recognition\\ Accuracy (\%)\end{tabular}} & \multicolumn{6}{c|}{AUC Score, Number of Gradients for Adversarial Attacks} \\ \cline{4-9} 
 & & & FGSM & DeepFool & BIM-a & BIM-b & JSMA & CWA \\ \hline
\multirow{2}{*}{ESC-10} & 8 & \textbf{85.67} & \textbf{0.9456}, 1 & \textbf{0.9310}, 429 & 0.9307, 612 & \textbf{0.9411}, 744 & \textbf{0.9324}, 781 & \textbf{0.9483}, 4205 \\ \cline{2-9} 
 & 16 & 82.04 & 0.9068, 1 & 0.9192, 672 & \textbf{0.9437}, 490 & 0.9347, 513 & 0.9018, 801 & 0.9216, 4439 \\ \hline \hline
\multirow{2}{*}{ESC-50} & 8 & 80.34 & \textbf{0.9462}, 1 & \textbf{0.9335}, 367 & 0.9161, 452 & 0.9314, 809 & 0.9168, 298 & 0.9233, 3981 \\ \cline{2-9} 
 & 16 & \textbf{85.97} & 0.9376, 1 & 0.9256, 409 & \textbf{0.9314}, 628 & \textbf{0.9419}, 701 & \textbf{0.9173}, 561 & \textbf{0.9236}, 4575 \\ \hline \hline
\multirow{2}{*}{UrbanSound8k} & 8 & \textbf{94.70} & \textbf{0.9401}, 1 & \textbf{0.9279}, 761 & \textbf{0.9315}, 841 & \textbf{0.9511}, 738 & \textbf{0.9207}, 691 & 0.9320, 4684 \\ \cline{2-9} 
 & 16 & 91.83 & 0.9321, 1 & 0.9274, 533 & 0.9125, 719 & 0.9408, 941 & 0.9139, 774 & \textbf{0.9430}, 4879 \\ \hline
\end{tabular}
\label{dwt_effect}
\end{table*}
Comparing the recognition accuracy of Tables~\ref{mfcc_sr_effect} and~\ref{stft_fft_effect} shows that STFT provides better discriminative features for the ResNet-18 classifier since such a model achieved lower recognition accuracy on MFCC representations. Additionally, while the AUC scores across the six attacks are not so different, ranging from 0.93 to 0.99, the number of gradient required for models trained on STFT spectrograms are considerably higher than MFCC spectrograms. In summary, STFT spectrograms provide a better accuracy and a little hard to attack, even if they can be fooled with high success by all six adversarial attacks.
%%%%%%%%%%%%%%%%%%%%%%%%%%%%%%%
\subsubsection{Adversarial Attacks for DWT representations}
%%%%%%%%%%%%%%%%%%%%%%%%%%%%%%%
There is no algorithmic approach for obtaining the optimal mother function to generate DWT spectrograms. Therefore, we have employed several functions, from simple Haar to complex Morlet to investigate the potential impacts on the recognition accuracy and on the adversarial robustness of the victim models. We exploited an analytical approach recasting multiple experiments. Table~\ref{mother_func_comp} shows that although complex Morlet mother function outperforms other mother functions in terms of recognition accuracy. However, it shows more vulnerability against adversarial examples, averaged over six attack algorithms with different budgets. 
\begin{table}[htpb!]
\centering
\caption{Comparison of mother functions on the performance of the models. Outperforming values are shown in bold face.}
\begin{tabular}{|c|c|c|c|}
\hline
Dataset & \begin{tabular}[c]{@{}c@{}}Mother\\ Function\end{tabular} & \begin{tabular}[c]{@{}c@{}}Average Recognition\\ Accuracy (\%)\end{tabular} & \begin{tabular}[c]{@{}c@{}}Average\\ AUC Score\end{tabular} \\ \hline
\multirow{3}{*}{ESC-10} & Haar & 82.14 & 95.14 \\ \cline{2-4} 
 & Mexican Hat & 84.51 & 94.19 \\ \cline{2-4} 
 & Complex Morlet & \textbf{85.67} & \textbf{95.61} \\ \hline \hline
\multirow{3}{*}{ESC-50} & Haar & 83.08 & 92.16 \\ \cline{2-4} 
 & Mexican Hat & 84.33 & 93.40 \\ \cline{2-4} 
 & Complex Morlet & \textbf{85.97} & \textbf{95.38} \\ \hline \hline
\multirow{3}{*}{UrbanSound8k} & Haar & 91.22 & 96.16 \\ \cline{2-4} 
 & Mexican Hat & 93.48 & 95.63 \\ \cline{2-4} 
 & Complex Morlet & \textbf{95.17} & \textbf{96.09} \\ \hline
\end{tabular}
\label{mother_func_comp}
\end{table}

Table~\ref{dwt_effect} compares the recognition accuracy of models trained on DWT representations with complex Morlet mother function. We have evaluated these models on DWT spectrograms with sampling rates of 8 kHz and 16 kHz. Whereas for ESC-50, sampling rate of 8 kHz shows better performance for the classifiers comparing their recognition accuracies. There are three findings in these tables. Firstly, averaged over all the allocated budgets for the attacks, models trained on DWT representations demonstrate a slightly higher robustness against adversarial attacks compared to MFCC and STFT spectrograms. Secondly, the highest recognition accuracy has been achieved for classifiers trained on DWT representations. Thirdly, the trade-off between recognition accuracy and adversarial robustness of the victim models are noticeable for different sampling rates. Moreover, the cost of the attack (number of gradient computations) for models trained on DWT is considerably higher than other two representations. 

\begin{figure}[htpb!]
  \centering
  \includegraphics[width=0.5\textwidth]{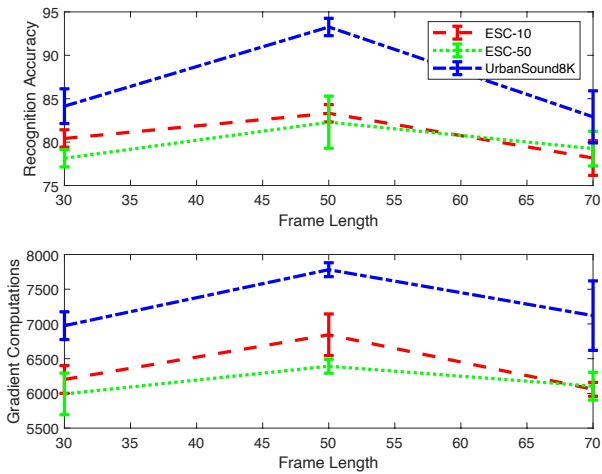}
  \caption{DWT frame length effect on the recognition accuracy and on the average cost of the attack for yielding AUC$>$0.9 over six adversarial algorithms for ResNet-18 models.}
  \label{mel-attack-pack5}
\end{figure}

\begin{figure*}[htb]
 \footnotesize 
\setlength{\tabcolsep}{2pt}
 \begin{tabular}{>{\centering}m{0.135\textwidth}
 >{\centering}m{0.135\textwidth}
 >{\centering}m{0.135\textwidth}
 >{\centering}m{0.135\textwidth}
 >{\centering}m{0.135\textwidth}
 >{\centering}m{0.135\textwidth}
 >{\centering\arraybackslash}m{0.135\textwidth}}
 {Original} & \multicolumn{6}{c}{Attacked Spectrograms} \\ \cmidrule(lr){2-7} 
{Spectrograms} & {FGSM} & {DeepFool} & {BIM-a} & {BIM-b} & {JSMA} & {CWA} \\
 \end{tabular}
\centering
 \subfloat[MFCC]{{\includegraphics[width=0.135\textwidth]{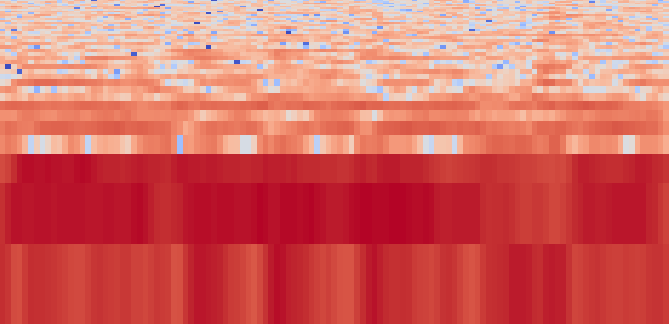} }}%
 \subfloat[$\left \| \delta \right \|_{2}=0.51, l{}'=2$]{{\includegraphics[width=0.135\textwidth]{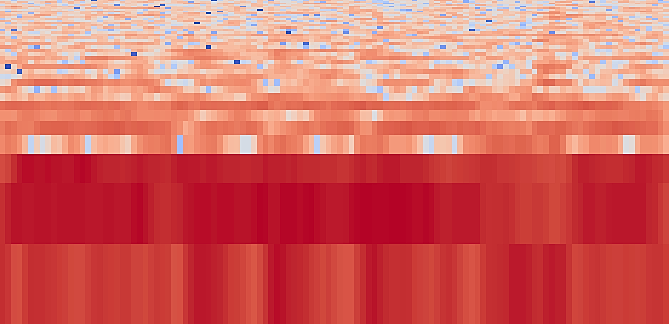} }}
 \subfloat[$\left \| \delta \right \|_{2}=0.67, l{}'=3$]{{\includegraphics[width=0.135\textwidth]{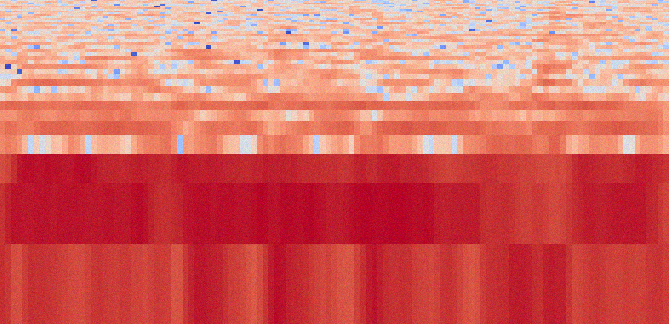} }}%
 \subfloat[$\left \| \delta \right \|_{2}=0.71, l{}'=4$]{{\includegraphics[width=0.135\textwidth]{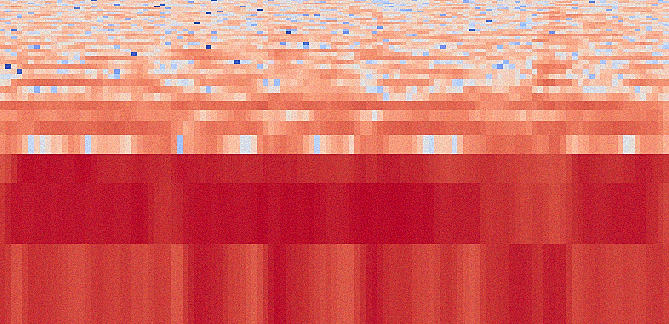} }}%
 \subfloat[$\left \| \delta \right \|_{2}=0.93, l{}'=5$]{{\includegraphics[width=0.135\textwidth]{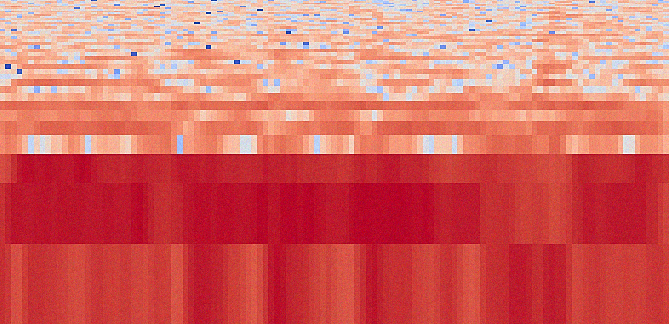} }}%
 \subfloat[$\left \| \delta \right \|_{0}=1.18, l{}'=6$]{{\includegraphics[width=0.135\textwidth]{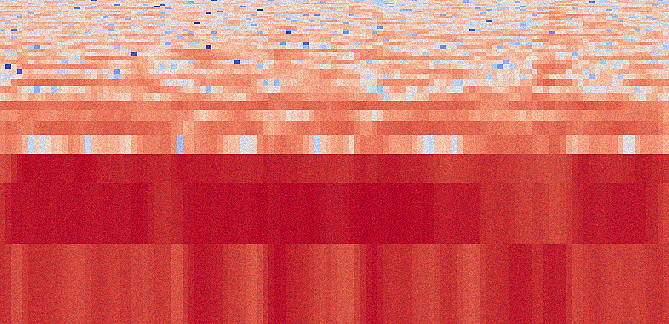} }}%
 \subfloat[$\left \| \delta \right \|_{2}=1.47, l{}'=7$]{{\includegraphics[width=0.135\textwidth]{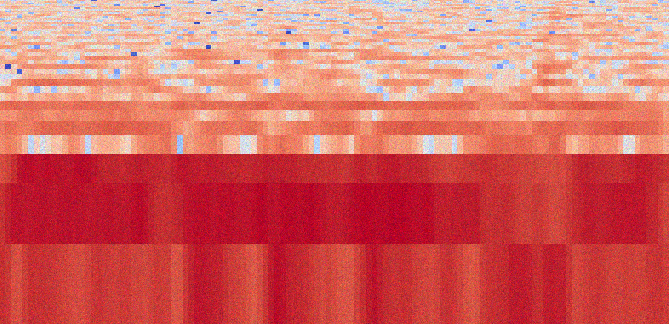} }}%
 \\
 \subfloat[STFT]{{\includegraphics[width=0.135\textwidth]{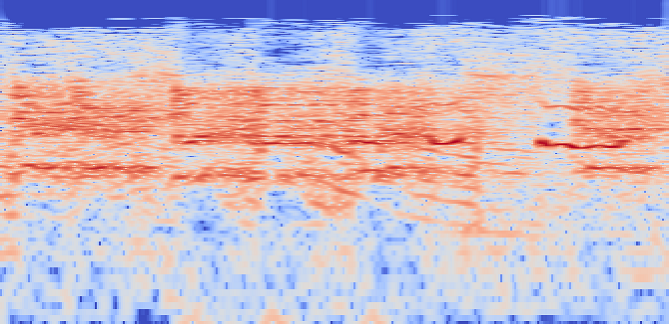} }}%
 \subfloat[$\left \| \delta \right \|_{2}=0.82, l{}'=2$]{{\includegraphics[width=0.135\textwidth]{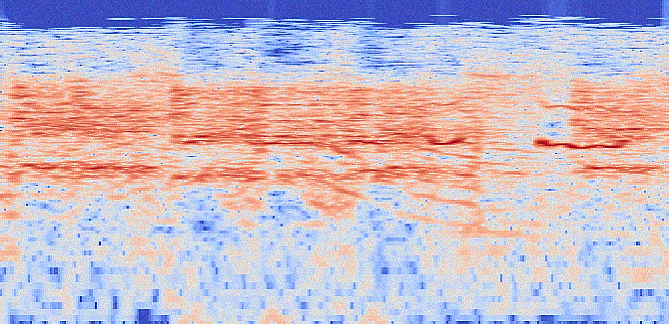} }}%
 \subfloat[$\left \| \delta \right \|_{2}=1.39, l{}'=3$]{{\includegraphics[width=0.135\textwidth]{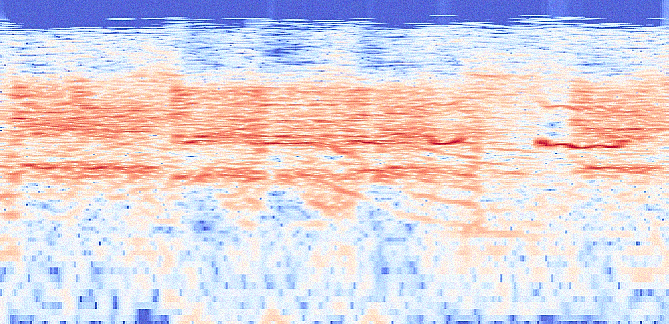} }}%
 \subfloat[$\left \| \delta \right \|_{2}=0.64, l{}'=4$]{{\includegraphics[width=0.135\textwidth]{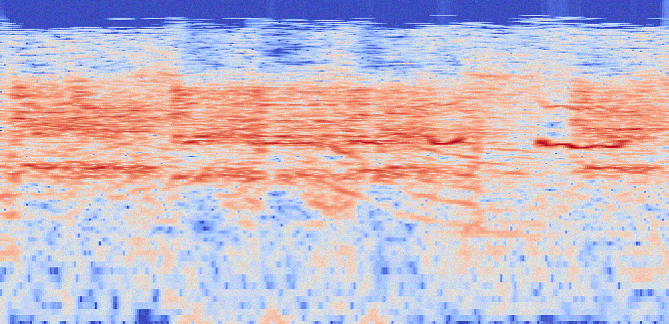} }}%
 \subfloat[$\left \| \delta \right \|_{2}=1.24, l{}'=5$]{{\includegraphics[width=0.135\textwidth]{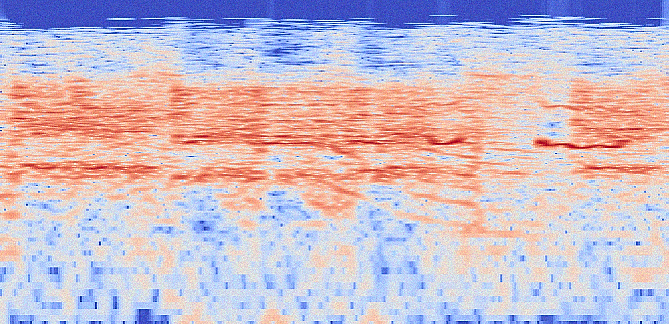} }}%
 \subfloat[$\left \| \delta \right \|_{0}=1.31, l{}'=6$]{{\includegraphics[width=0.135\textwidth]{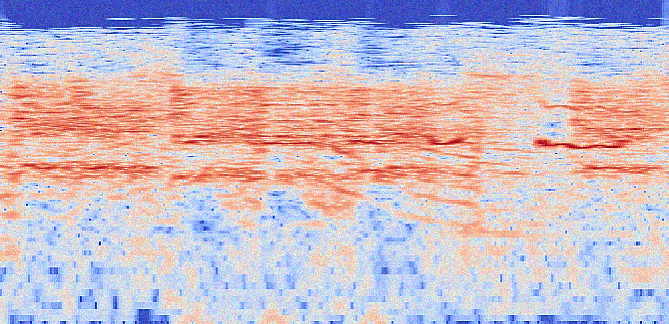} }}%
 \subfloat[$\left \| \delta \right \|_{2}=1.73, l{}'=7$]{{\includegraphics[width=0.135\textwidth]{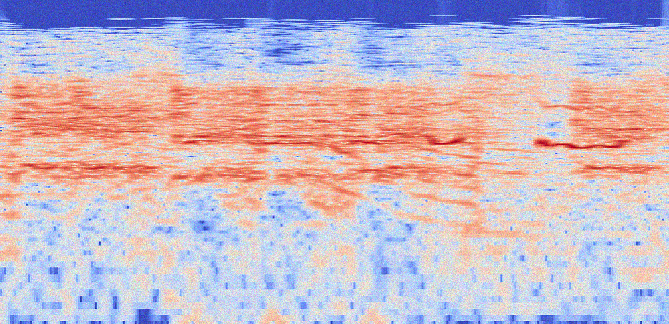} }}%
 \\
 \subfloat[DWT]{{\includegraphics[width=0.135\textwidth]{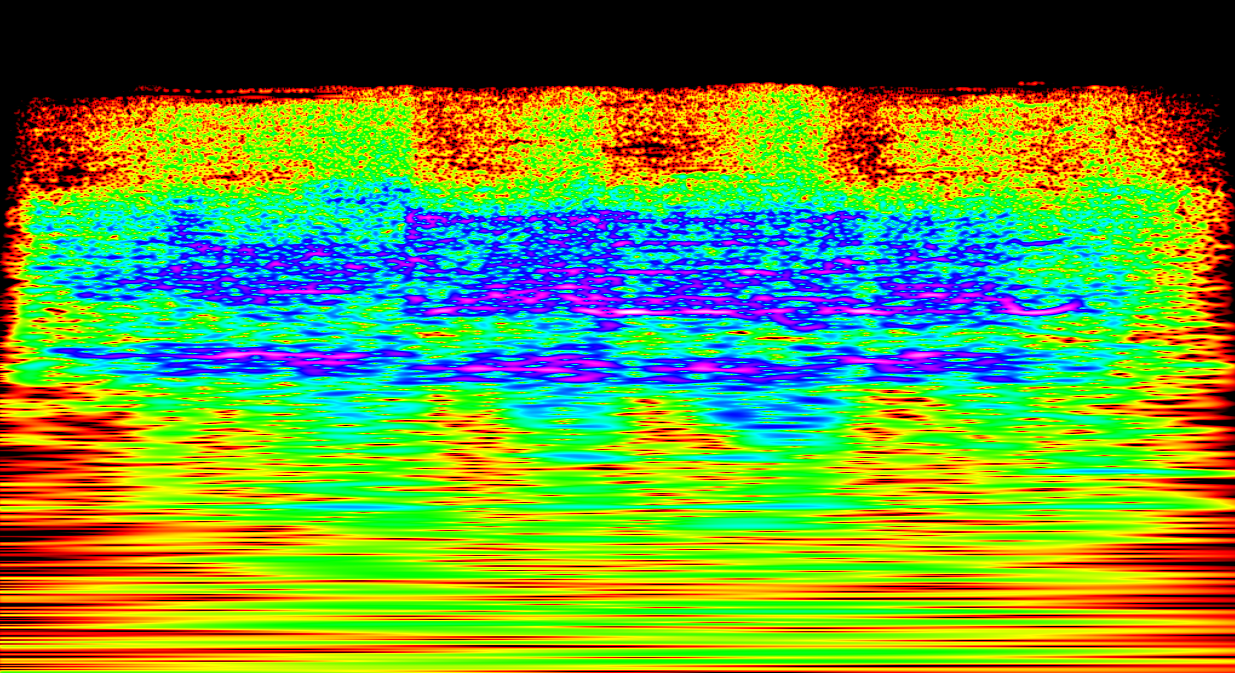} }}%
 \subfloat[$\left \| \delta \right \|_{2}=1.13, l{}'=2$]{{\includegraphics[width=0.135\textwidth]{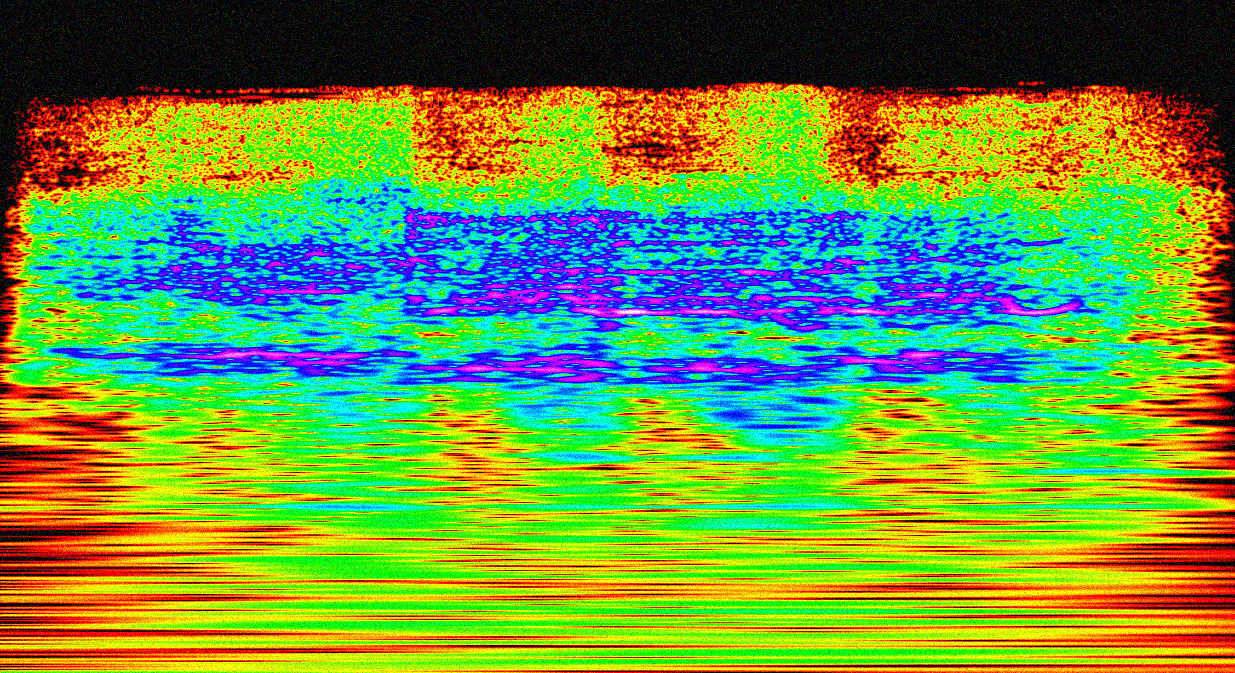} }}%
 \subfloat[$\left \| \delta \right \|_{2}=1.36, l{}'=3$]{{\includegraphics[width=0.135\textwidth]{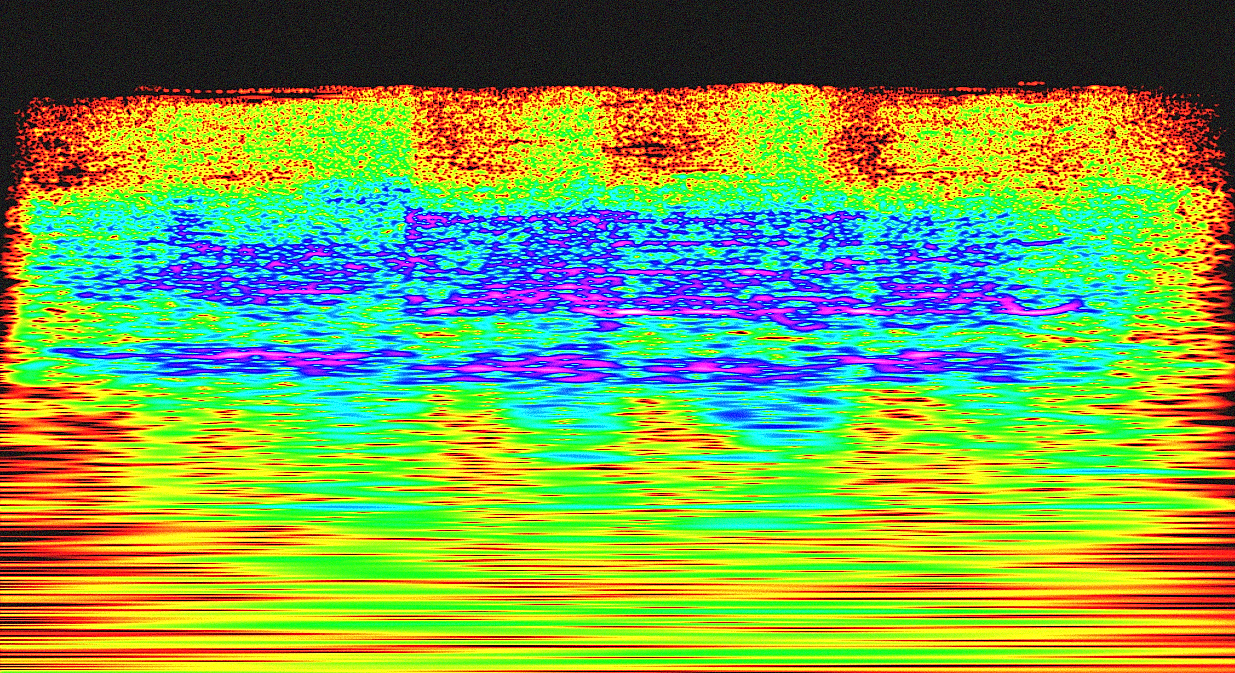} }}%
 \subfloat[$\left \| \delta \right \|_{2}=1.96, l{}'=4$]{{\includegraphics[width=0.135\textwidth]{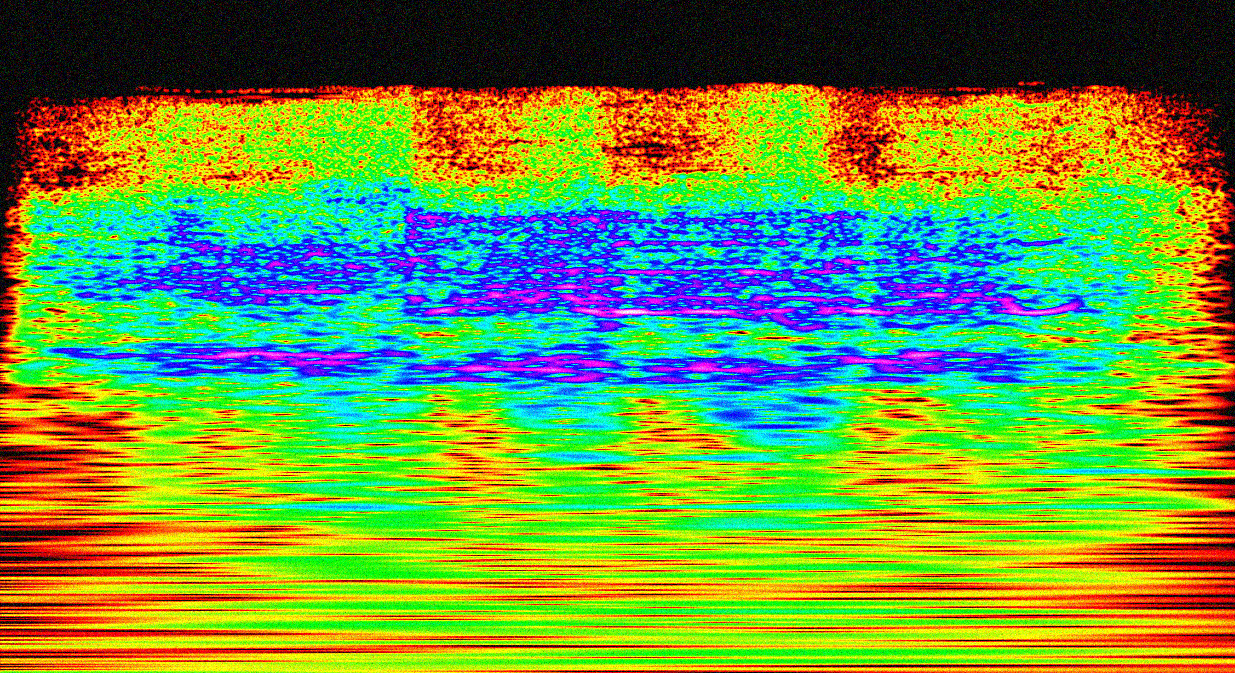} }}%
 \subfloat[$\left \| \delta \right \|_{2}=1.49, l{}'=5$]{{\includegraphics[width=0.135\textwidth]{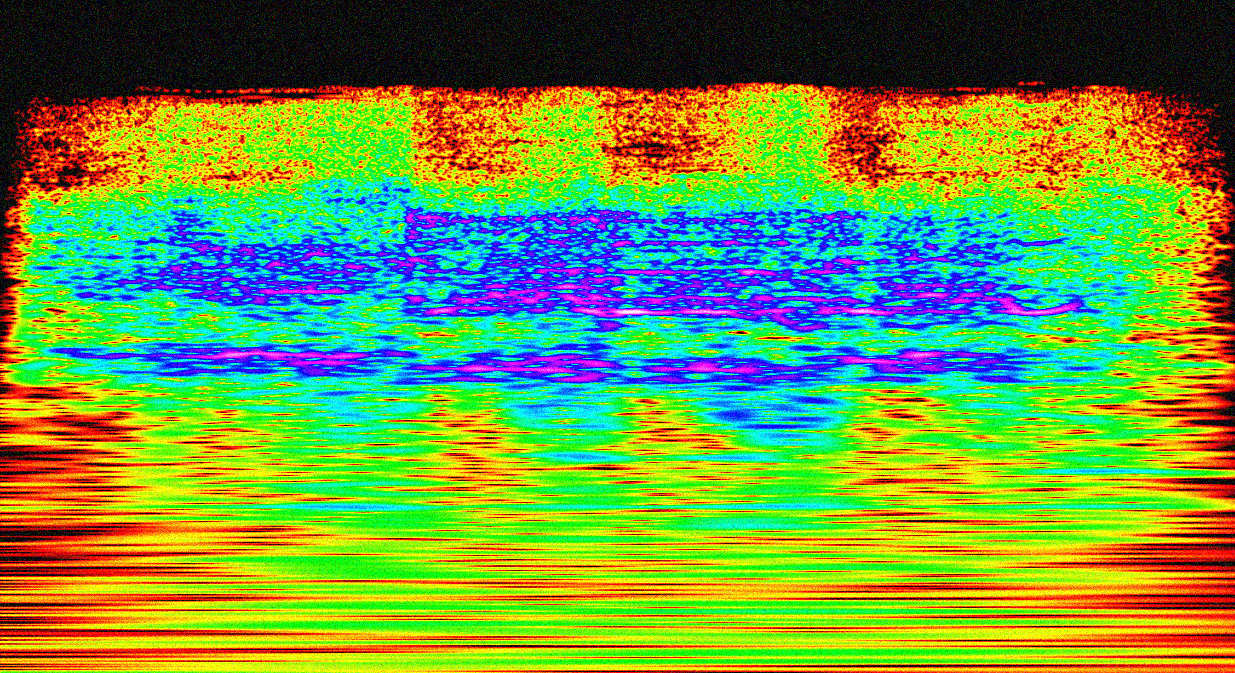} }}%
 \subfloat[$\left \| \delta \right \|_{0}=2.03, l{}'=6$]{{\includegraphics[width=0.135\textwidth]{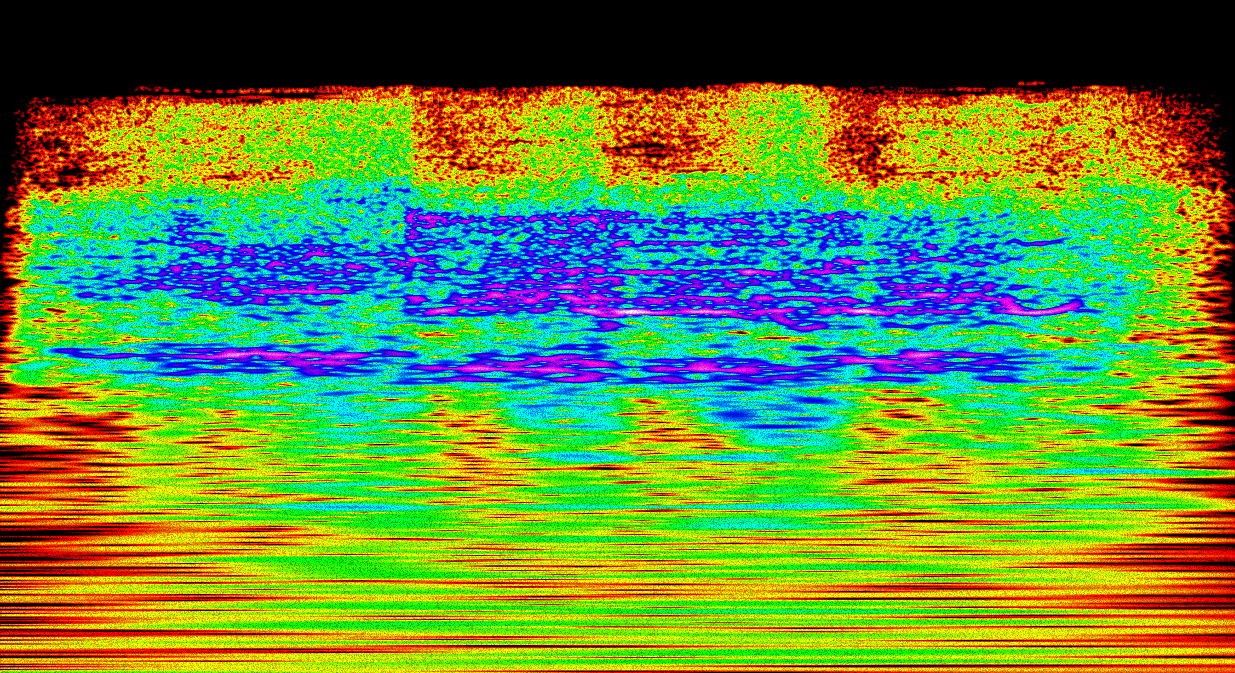} }}%
 \subfloat[$\left \| \delta \right \|_{2}=2.38, l{}'=7$]{{\includegraphics[width=0.135\textwidth]{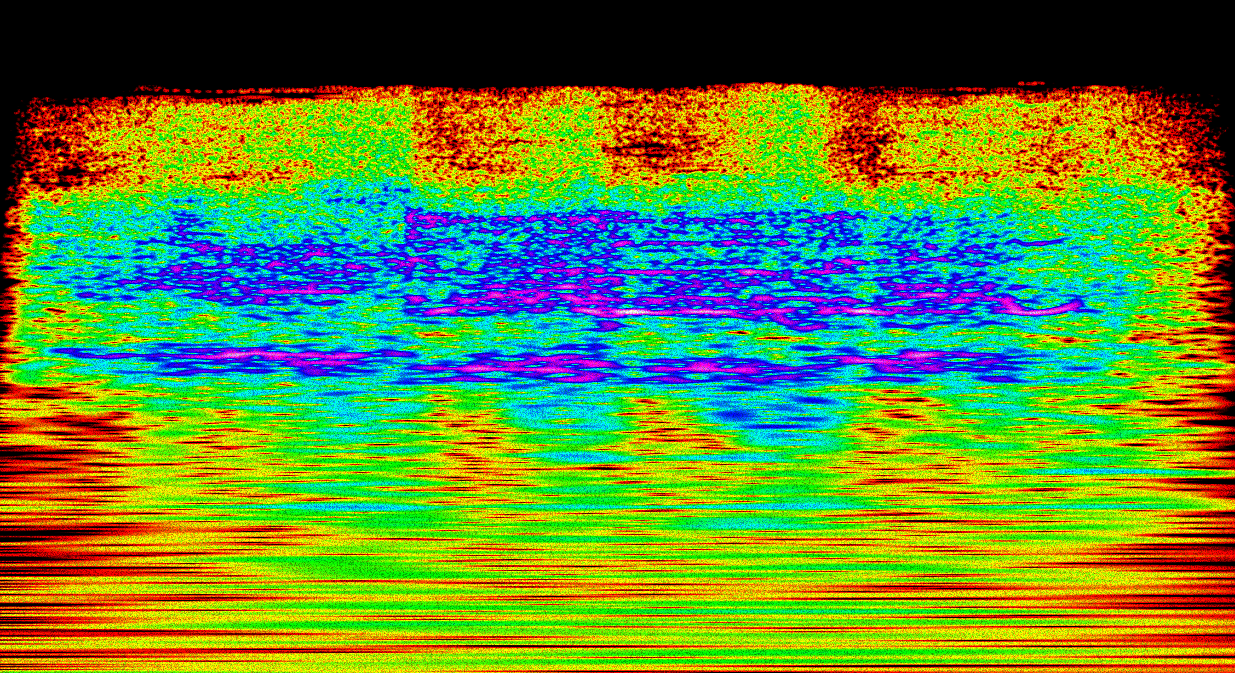} }}%
\caption{Crafted adversarial spectrograms for the three audio representations. The original audio sample has been randomly selected from the class of dog bark ($l=1$). Examples shown in columns two to seven are associated with the six adversarial attacks for the original input sample. Required perturbation ($\delta$) and the target labels ($l{}'$) are shown under each spectrogram.}
\label{attack_spec_compr}
\end{figure*}

In all these experiments, we assumed a frame length of 50 ms with 50\% overlapping to convolve the input signal with mother functions. We have also carried out experiments on studying the potential effect of frame length in performance of the models. They showed that short frame lengths (e.g., 30 ms) drop the recognition performance of the models for the three benchmarking datasets. Additionally, long frames such as 50 ms introduce a high redundancy in frequency plots, which results in dropping the recognition accuracy (see Fig.~\ref{mel-attack-pack5}). Fig.~\ref{attack_spec_compr} visually compares crafted adversarial examples for the three representations. Although they are visually very similar to their legitimate counterparts, they confidently drive the classifier toward wrong predictions. This showcases the active threat of adversarial attacks for the sound recognition models. 

\captionsetup[subfigure]{labelformat=empty}

%%%%%%%%%%%%%%%%%%%%%%%%%%%%%%%
\section{Discussion}
\label{discuss:sec}
%%%%%%%%%%%%%%%%%%%%%%%%%%%%%%%
In this section we provide additional discussion regarding our results. We briefly discuss some secondary aspects of our experiments which could be relevant for future studies. 

%%%%%%%%%%%%%%%%%%%%%%%%%%%%%%
\subsection{Neural Network Architecture} %\noindent \textbf{Neural network architecture.}
%%%%%%%%%%%%%%%%%%%%%%%%%%%%%%%
For selecting the front-end classifier, we measured recognition accuracy and total number of training parameters for all candidates. We explored DL architectures without residual blocks (AlexNet) and with inception blocks (GoogLeNet) for the choice of victim classifiers. Our experiments unveiled that these dense networks do not outperform ResNet-18 in terms of recognition accuracy. Although average recognition accuracy of ResNet-18 and GoogLeNet are competitive on spectrograms, the latter has 1.41$\times$ more training parameters. On average, recognition performance of AlexNet is 8\% lower than ResNet-18, even if it has 61\% fewer parameters. Furthermore, recognition performance of other ResNet models such as ResNet-34 and ResNet-56 are very competitive to ResNet-18, but the latter requires 50\% fewer parameters. In comparing the robustness of these models against adversarial attacks, they all can reach fooling rates higher than 95\%. Taking the allocated budgets into account, the ResNet-18 is the costliest network in terms of the number of required gradient computations for the adversary, followed by GoogLeNet and AlexNet. 
% Stopped here 27-7

%%%%%%%%%%%%%%%%%%%%%%%%%%%%%%%
\subsection{Data Augmentation} %\noindent \textbf{Data augmentation.}
%%%%%%%%%%%%%%%%%%%%%%%%%%%%%%%
For improving the performance of the classifiers, we augmented the original datasets only at waveform level (1D) using time-stretching filter except for DWT representations which we additionally scaled the spectrograms by a logarithmic function. Removing 1D data augmentation negatively affects recognition accuracy of the models with drop ratios of about 0.056\%, 0.036\%, and 0.029\% for MFCC, STFT, and DWT spectrograms, respectively. For measuring the robustness of these models against adversarial examples, we run attack algorithms on random batches of size 100 among the entire datasets. The experimental results have shown that for reaching the fooling rates as close as the values reported in Tables~\ref{mfcc_sr_effect} to \ref{dwt_effect}, less gradient computation is required mainly for JSMA and CWA attacks.

%%%%%%%%%%%%%%%%%%%%%%%%%%%%%%%
\subsection{Adversarial on Raw Audio} %\noindent \textbf{Adversarial on Raw Audio.}
%%%%%%%%%%%%%%%%%%%%%%%%%%%%%%%
Optimizing Eq.~\ref{general_adv_formula} even for a very short 1D audio signal sampled at a low rate is very costly and they are not transferable while being played over the air \cite{carlini2018audio}. Toward addressing this interesting open problem, we trained several end-to-end ConvNets on randomly selected batches of environmental sound datasets. Upon running both targeted and non-targeted attacks against ConvNets we could reduce performance of victim classifiers by 30\% in average. Interestingly, multiplying the adversarial examples by a random small scalar restore correct label of the audio waveforms. In other words, whereas adversarial spectrograms, 1D adversarial audio waveforms are not resilient against any additional perturbation.

%%%%%%%%%%%%%%%%%%%%%%%%%%%%%%%
\subsection{Adversarial Transferability} %\noindent \textbf{Adversarial transferability.}
%%%%%%%%%%%%%%%%%%%%%%%%%%%%%%%
Transferability of adversarial examples not only is dependent to the classifier, but also to audio representations. We investigate this on deep neural networks trained on different spectrograms. Table~\ref{table_transf} reports the transferability ratios averaged over budgets with batch sizes of 100. Crafted adversarial examples for victim models are less transferable in MFCC representations while DWT spectrograms have higher transferring rates on average. Examples generated in STFT domain are more transferable compared to MFCC, this may be due to the higher order of information in STFT spectrograms.

\begin{table*}[htbp]
\centering
\caption{Average transferability ratio of adversarial examples among ConvNets. Higher ratios are shown in boldface.}
\begin{tabular}{|c|c||c|c|c||c|c|c||c|c|c|}
\hline
 & & \multicolumn{3}{c||}{MFCC} & \multicolumn{3}{c||}{STFT} & \multicolumn{3}{c|}{DWT} \\ \cline{3-11} 
\multirow{-2}{*}{Dataset} & \multirow{-2}{*}{Models} & ResNet-18 & GoogLeNet & AlexNet & ResNet-18 & GoogLeNet & AlexNet & ResNet-18 & GoogLeNet & AlexNet \\ \hline
 & ResNet-18 & \cellcolor[HTML]{9B9B9B}{\color[HTML]{333333} 1} & \textbf{0.672} & 0.568 & \cellcolor[HTML]{9B9B9B}1 & \textbf{0.713} & 0.641 & \cellcolor[HTML]{9B9B9B}1 & 0.761 & \textbf{0.774} \\ \cline{2-11} 
 & GoogLeNet & \textbf{0.693} & \cellcolor[HTML]{9B9B9B}1 & 0.480 & \textbf{0.637} & \cellcolor[HTML]{9B9B9B}1 & 0.519 & 0.646 & \cellcolor[HTML]{9B9B9B}1 & \textbf{0.684} \\ \cline{2-11} 
\multirow{-3}{*}{ESC-10} & AlexNet & 0.491 & \textbf{0.521} & \cellcolor[HTML]{9B9B9B}1 & 0.540 & \textbf{0.562} & \cellcolor[HTML]{9B9B9B}1 & 0.633 & \textbf{0.701} & \cellcolor[HTML]{9B9B9B}1 \\ \hline \hline
 & ResNet-18 & \cellcolor[HTML]{9B9B9B}1 & \textbf{0.644} & 0.519 & \cellcolor[HTML]{9B9B9B}1 & \textbf{0.661} & 0.609 & \cellcolor[HTML]{9B9B9B}1 & \textbf{0.755} & 0.732 \\ \cline{2-11} 
 & GoogLeNet & \textbf{0.630} & \cellcolor[HTML]{9B9B9B}1 & 0.531 & \textbf{0.578} & \cellcolor[HTML]{9B9B9B}1 & 0.569 & 0.507 & \cellcolor[HTML]{9B9B9B}1 & \textbf{0.676} \\ \cline{2-11} 
\multirow{-3}{*}{ESC-50} & AlexNet & 0.523 & \textbf{0.536} & \cellcolor[HTML]{9B9B9B}1 & 0.551 & \textbf{0.601} & \cellcolor[HTML]{9B9B9B}1 & 0.614 & \textbf{0.699} & \cellcolor[HTML]{9B9B9B}1 \\ \hline \hline
 & ResNet-18 & \cellcolor[HTML]{9B9B9B}1 & 0.627 & \textbf{0.677} & \cellcolor[HTML]{9B9B9B}1 & 0.611 & \textbf{0.710} & \cellcolor[HTML]{9B9B9B}1 & \textbf{0.714} & 0.713 \\ \cline{2-11} 
 & GoogLeNet & \textbf{0.634} & \cellcolor[HTML]{9B9B9B}1 & 0.503 & 0.563 & \cellcolor[HTML]{9B9B9B}1 & \textbf{0.699} & \textbf{0.723} & \cellcolor[HTML]{9B9B9B}1 & 0.707 \\ \cline{2-11} 
\multirow{-3}{*}{Urban8k} & AlexNet & 0.577 & \textbf{0.583} & \cellcolor[HTML]{9B9B9B}1 & 0.703 & \textbf{0.735} & \cellcolor[HTML]{9B9B9B}1 & 0.705 & \textbf{0.678} & \cellcolor[HTML]{9B9B9B}1 \\ \hline
\end{tabular}
\label{table_transf}
\end{table*}

%%%%%%%%%%%%%%%%%%%%%%%%%%%%%%%
\section{Conclusion}
\label{sec:con}
%%%%%%%%%%%%%%%%%%%%%%%%%%%%%%%
In this paper, we demonstrated the inverse relationship between recognition accuracy and robustness of ResNet-18 trained on 2D representations of environmental audio signals averaged over allocated budgets by the adversary. This relation is generalizable to other DL architectures and this is a common behavior for models trained on spectrograms.
Additionally, we showed that our front-end classifier can reach the highest recognition accuracy when it is trained on DWT representation. Furthermore, attacking this model is on average more costly for the adversary compared to models trained on MFCC and STFT representations. This proves the superiority of DWT representation for environmental sound recognition. Moreover, we examined the transferability of crafted adversarial examples among AlexNet, GoogLeNet, and ResNet-18 for the three spectrogram representations. According to our results, the lowest transferability ratio was achieved for MFCC spectrograms averaged over six different adversarial attacks. In our future studies, we are determined to investigate this property for networks trained on speech datasets.

\bibliographystyle{IEEEtran}
\bibliography{IEEEabrv,mybib}

\vfill


\end{document}